\documentclass[acmsmall]{acmart}

\usepackage{metalogo}
\usepackage{smartdiagram}
\usepackage{forest}
\usepackage{cleveref}
\usepackage{booktabs}
\usepackage{multirow}
\usepackage{makecell}
\usepackage{subfigure}
\usepackage[misc]{ifsym} 
\usepackage{balance}
\usepackage{listings}
\usepackage{color, xcolor}
\usepackage{tcolorbox}
\usepackage{amsmath}
\usepackage{algorithm}
\usepackage{graphicx}
\usepackage{algorithmicx}
\usepackage{algpseudocode}
\usepackage{textcomp}
\usepackage{diagbox}
\usepackage{colortbl}
\usepackage{seqsplit}
\usepackage{xurl}
\usepackage{url}

\acmJournal{TOSEM}
\acmYear{2026} \acmVolume{1} \acmNumber{1} \acmArticle{}
\acmMonth{1} \acmPrice{} \acmDOI{10.1145/3789262}

\begin{document}

\title{ThinkFL: Self-Refining Failure Localization for Microservice Systems via Reinforcement Fine-Tuning}

\author{Lingzhe Zhang$^{\dag}$}
\affiliation{%
	\institution{Peking University}
	\city{Beijing}
	\country{China}}
\email{zhang.lingzhe@stu.pku.edu.cn}

\author{Yunpeng Zhai$^{\dag}$}
\thanks{$^{\dag}$Equal contribution}
\affiliation{%
	\institution{Alibaba Group}
	\country{China}}
\email{zhaiyunpeng.zyp@alibaba-inc.com}

\author{Tong Jia$^{\ast}$}
\thanks{$^{\ast}$Corresponding author}
\affiliation{%
	\institution{Peking University; National Key Laboratory of Data Space Technology and System}
	\city{Beijing}
	\country{China}}
\email{jia.tong@pku.edu.cn}

\author{Chiming Duan}
\affiliation{%
	\institution{Peking University}
	\city{Beijing}
	\country{China}}
\email{duanchiming@stu.pku.edu.cn}

\author{Siyu Yu}
\affiliation{%
	\institution{Peking University}
	\city{Beijing}
	\country{China}}
\email{gaiusyu6@gmail.com}

\author{Jinyang Gao}
\affiliation{%
	\institution{Alibaba Group}
	\country{China}}
\email{jinyang.gjy@alibaba-inc.com}

\author{Bolin Ding}
\affiliation{%
	\institution{Alibaba Group}
	\country{United States}}
\email{bolin.ding@alibaba-inc.com}

\author{Zhonghai Wu}
\affiliation{%
	\institution{Peking University}
	\city{Beijing}
	\country{China}}
\email{wuzh@pku.edu.cn}

\author{Ying Li$^{\ast}$}
\affiliation{%
	\institution{Peking University}
	\city{Beijing}
	\country{China}}
\email{li.ying@pku.edu.cn}

\renewcommand{\shortauthors}{Lingzhe Zhang et al.}

\begin{abstract}
	As modern microservice systems grow increasingly popular and complex—often consisting of hundreds or even thousands of interdependent components—they are becoming more susceptible to frequent and subtle failures. Ensuring system reliability therefore hinges on accurate and efficient failure localization. Traditional failure localization approaches based on small models lack the flexibility to adapt to diverse failure scenarios, while recent LLM-based methods suffer from two major limitations: they often rely on rigid invocation workflows that constrain the model’s ability to dynamically explore effective localization paths, and they require resource-intensive inference, making them cost-prohibitive for real-world deployment. To address these challenges, we explore the use of reinforcement fine-tuning to equip lightweight LLMs with reasoning and self-refinement capabilities, significantly improving the cost-effectiveness and adaptability of LLM-based failure localization. We begin with an empirical study to identify three key capabilities essential for accurate localization. Building on these insights, we propose a progressive multi-stage GRPO fine-tuning framework, which integrates a multi-factor failure localization grader and a recursion-of-thought actor module. The resulting model, \textbf{ThinkFL}, not only outperforms existing state-of-the-art LLMs and baseline methods in localization accuracy but also reduces end-to-end localization latency from minutes to seconds, demonstrating strong potential for real-world applications.
\end{abstract}

\begin{CCSXML}
	<ccs2012>
	<concept>
	<concept_id>10011007.10011074.10011111.10011696</concept_id>
	<concept_desc>Software and its engineering~Maintaining software</concept_desc>
	<concept_significance>500</concept_significance>
	</concept>
	</ccs2012>
\end{CCSXML}

\ccsdesc[500]{Software and its engineering~Maintaining software}

\keywords{Failure Localization, Trace, Large Language Model, Reinforcement Fine-Tuning}

%\received{20 June 2024}
%\received[revised]{}
%\received[accepted]{}

\maketitle

\section{Introduction}

Modern microservice architectures have grown increasingly intricate due to evolving runtime conditions and complex inter-service interactions~\cite{zhang2025survey}. These systems typically comprise hundreds or even thousands of tightly coupled subsystems, where a fault in any single component can cascade into system-wide performance degradations~\cite{waseem2021design}. Consequently, timely and accurate identification of the underlying cause of such failures is essential to maintaining overall system reliability~\cite{xu2025openrca, wang2024mrca}.

Nevertheless, pinpointing the exact root cause remains a formidable challenge due to the complex interdependencies among subsystems~\cite{sun2025interpretable}. Each request typically triggers an elaborate chain of invocations that spans multiple components—ranging from services and service instances to physical hosts—and encompasses various interactions, such as Application Programming Interface (API) calls, database queries, and other inter-component communications. The inherent dynamism and heterogeneity of these interactions render the task of isolating the true root cause both complex and non-trivial.

To enable root cause localization, extensive research has been conducted in this area. Many studies have leveraged machine learning and deep learning models to address this challenge. Microscope~\cite{lin2018microscope} constructs causality graphs and employs a depth-first search strategy to detect front-end anomalies. CIRCA~\cite{li2022causal} builds a causal Bayesian network using regression-based hypothesis testing and descendant adjustment to infer faulty components. RUN~\cite{lin2024root} integrates time series forecasting with neural Granger causal discovery and a personalized PageRank algorithm to efficiently recommend the top-k root causes. MicroRank~\cite{yu2021microrank} constructs a trace coverage tree to capture dependencies between requests and service instances, applying the PageRank algorithm to rank potential root causes. TraceRank~\cite{yu2023tracerank} combines spectrum analysis with a PageRank-based random walk to pinpoint abnormal services. CRISP~\cite{zhang2022crisp} performs critical path analysis to drill down into latency issues, while TraceConstract~\cite{zhang2024trace} employs sequence representations, contrast sequential pattern mining, and spectrum analysis to efficiently localize multi-dimensional root causes.

However, these approaches rely on predefined service causal graphs, fixed statistical models of fault-symptom relationships, or learned associations from training data, fundamentally operating as static, one-time analyses. Furthermore, deep learning-based methods often struggle to provide interpretable reasoning paths, which are crucial for Site Reliability Engineers (SREs) to validate diagnoses and implement effective resolutions~\cite{beyer2018site, yao2024chain, yu2023nezha, zhang2025adaptive}. Fortunately, large language models (LLMs) offer promising solutions to these challenges, leading to the emergence of LLM-based approaches. For instance, mABC~\cite{zhang2024mabc} introduces a multi-agent, blockchain-inspired collaboration framework, where multiple LLM-based agents follow a structured workflow and coordinate via blockchain-inspired voting mechanisms. RCAgent~\cite{wang2024rcagent} utilizes a tool-augmented LLM framework that incorporates log and code data to perform root cause analysis tasks, such as predicting root causes, identifying solutions, gathering evidence, and determining responsibilities.

Although these LLM-based methods have significantly advanced failure localization, several challenges remain in practical deployment:

\begin{itemize}
	\item \textbf{Rigid Invocation Workflows.} Current LLM-based approaches typically rely on fixed invocation workflows—for instance, multi-agent frameworks use a set of pre-defined agents that are called in a predetermined sequence~\cite{zhang2024mabc, wang2024rcagent, pei2025flow, li2025coca, wang2025tamo, ren2025multi}. For example, mABC~\cite{zhang2024mabc} adopts a staged multi-agent pipeline in which alerts flow through predefined roles in order—Alert Receiver (A1) $\rightarrow$ Process Scheduler (A2) $\rightarrow$ Data Detective (A3) / Dependency Explorer (A4) / Probability Oracle (A5) / Fault Mapper (A6) $\rightarrow$ Solution Engineer (A7)—with each agent bound to a specific diagnostic subtask. Similarly, Flow-of-Action~\cite{pei2025flow} constrains the agent’s behavior using Standard Operating Procedures (SOPs): the model must follow an SOP-guided sequence of actions. This inflexibility limits the LLM’s ability to dynamically explore and determine the correct failure localization path under varying conditions.
	\item \textbf{Resource-Intensive Inference.} The models employed in these approaches are generally ultra-large (with parameter counts $\geq$ 70 billion or more) or are closed-source, which restricts their practicality in terms of inference speed and computational cost. For instance, mABC~\cite{zhang2024mabc}, Flow-of-Action~\cite{pei2025flow}, and TAMO~\cite{wang2025tamo} report their strongest performance only when powered by closed-source frontier models such as GPT-4-Turbo. Likewise, COCA~\cite{li2025coca} relies on GPT-3.5, GPT-4o, Llama-3.1-405B, Gemini-1.5-Pro, and Claude-3.5-Sonnet to achieve competitive results. More recently, KnowledgeMind~\cite{ren2025multi} demonstrates the need for ultra-large open-source models (e.g., LLaMA-3.3-70B, Qwen-QwQ) or GPT-4-Turbo to realize its full potential. While such settings yield strong benchmarks, they highlight a critical challenge: in real-world failure localization scenarios, even a mature small-scale microservice system (with fewer than 10 nodes) may encounter issues every few minutes, making rapid and cost-effective problem resolution essential to minimize impact and loss.
\end{itemize}

Recognizing these limitations, we require a model that is both computationally efficient and capable of flexible, autonomous invocation. To address the first requirement, we leverage lightweight LLMs, which reduce deployment and inference costs while providing general-purpose semantic understanding that can be further specialized~\cite{dubey2024llama, yang2025qwen3}. To achieve flexible and autonomous invocation, the model must be able to self-explore multi-step diagnostic paths and optimize action selection. Reinforcement fine-tuning provides a principled mechanism for this: it enables the model to learn policies that select the effective sequence of actions based on feedback~\cite{rafailov2023direct, schulman2017proximal, shao2024deepseekmath}. Therefore, employing reinforcement fine-tuning to adapt lightweight LLMs (with parameter counts $\textless$ 10 billion) naturally addresses both constraints, allowing the model to iteratively explore effective failure localization pathways while avoiding the limitations of rigid invocation workflows and resource-intensive inference.

Based on this idea, we first conduct an empirical study to explore how to equip lightweight LLMs with failure localization capabilities. Our study identifies three core capabilities essential for effective failure localization: (i) a high dependence on standardized tool invocation, (ii) the need for exploratory behavior under standardized tool usage, and (iii) an integrated reward mechanism that accounts for both the reasoning process and the final output.

Based on our empirical study findings, we introduce ThinkFL, a self-refining failure localization model for microservice systems developed via reinforcement fine-tuning. ThinkFL leverages a Recursion-of-Thought Actor that integrates multiple data tools—combining trace analysis and metrics evaluation—to iteratively generate a ranked list of potential root causes along with a fully interpretable inference path. This actor output is then validated by expert SREs and further refined using a Multi-Factor Failure Localization Grader, which merges the model’s predictions with ground-truth feedback to compute a comprehensive score.

The entire fine-tuning process is driven by our Progressive Multi-Stage Group Relative Policy Optimization (GRPO) Fine-Tuning approach, which comprises three distinct stages: a format-aware policy priming stage to ensure the model reliably invokes tools in a standardized manner, a guided exploration augmentation stage to enhance the model’s capacity for extensive exploratory search, and a rank-oriented refinement stage to improve the final ranking and accuracy of the identified root cause.

Our experiments demonstrate that ThinkFL not only significantly outperforms existing large-scale LLMs in failure localization but also surpasses state-of-the-art root cause localization methods. In addition, ThinkFL achieves over 10× faster inference speed compared to these large models. By applying the ThinkFL training methodology to fine-tune various lightweight LLMs, we further validate the effectiveness and generalizability of our reinforcement learning approach. In summary, the key contributions of this work are as follows:

\begin{itemize} 
	\item We conduct a comprehensive study to explore how lightweight LLMs can be endowed with failure localization capabilities. Our study identifies three core capabilities essential for effective LLM-based failure localization. 
	\item Drawing inspiration from these insights, we propose ThinkFL, a self-refining failure localization model for microservice systems, achieved through reinforcement fine-tuning. ThinkFL introduces a novel progressive multi-stage GRPO fine-tuning approach, addressing the challenges identified in our empirical study while minimizing inference cost and iteratively exploring effective failure localization pathway. 
	\item We evaluate ThinkFL on six diverse datasets, demonstrating its superior performance. Experimental results reveal that ThinkFL achieves both higher accuracy and faster inference speed compared to large-scale baseline models and state-of-the-art failure localization methods. 
\end{itemize}

\section{Background \& Motivation}

This section outlines the essential background underpinning our approach, including what is a failure in microservice systems, a formal description of the failure localization (FL) problem, the role of distributed tracing as a core data modality in system observability, and the emerging use of reinforcement fine-tuning (RFT) to enhance the reasoning and adaptability of LLMs.

\subsection{Failure in Microservice Systems}

In large-scale microservice systems, failures typically manifest as abnormal behaviors observable at the request or application layer. For example, a user request may experience significantly increased latency, return an unexpected error code, or fail to complete altogether. At the system level, such failures are reflected in observability signals, including anomalous spikes in service metrics (e.g., error rates, CPU usage, or memory consumption), suspicious log patterns (e.g., repeated timeout errors), or incomplete traces where downstream calls do not return. These symptoms provide indirect evidence of the underlying faulty component but rarely reveal the exact root cause directly.

Specifically, in this work, a \emph{failure} is defined as an \textbf{abnormal episode} detectable from request-level observations within a short time window (typically one minute). An episode is triggered when any of the following request-observable signals cross predefined thresholds:
\begin{itemize}
	\item \textbf{Correctness/availability}: a surge in HTTP $5xx$ error rate, request timeout rate, or sudden unavailability of a service;
	\item \textbf{Performance}: tail-latency (e.g., $p95/p99$) exceeding the SLO budget, or throughput dropping sharply under steady load.
\end{itemize}

We note that resource saturation metrics (e.g., CPU, memory, GC pauses, I/O wait, or connection pool exhaustion) are excluded as primary failure categories because they are not directly observable from request-level signals. Instead, their impact typically manifests as correctness or performance degradations, which are captured by the above observables.

These conditions are reflected in observability signals—metrics, logs, and distributed traces—and grouped into a single failure \emph{episode} by temporal proximity and correlation.

We focus on application-layer \emph{correctness} and \emph{performance} failures in microservice systems, explicitly \underline{excluding} security incidents, user-intent/query-quality issues, front-end-only rendering glitches, and source-level bug localization. Our target is \emph{component-level} localization, identifying faulty entities such as services, pods, or hosts.

\subsection{Failure Localization}

Failure localization is a central problem in system failure diagnosis, aiming to identify the specific component—such as a microservice, process, or host—responsible for abnormal behavior. It is often considered the second stage of failure diagnosis, following coarse-grained failure category classification (e.g., identifying a CPU overload versus a network delay). Accurate localization is essential for reducing system recovery time and minimizing service disruption.

In this work, we define the \emph{root cause} as the system component directly responsible for the observed failure, with granularity at multiple levels: service, pod, and node. At the service level, the root cause corresponds to the microservice exhibiting the abnormal behavior, such as \texttt{CheckoutService}. At the pod level, the root cause can be a specific instance of a service (e.g., \texttt{checkoutservice-0}) that is malfunctioning. At the node level, the root cause is the physical or virtual machine (e.g., \texttt{node-0}) hosting the affected components.

For each failure episode, the root cause occurs at exactly one of these levels. Precise identification at the correct level allows operators to apply targeted remediation strategies. For example, if a request experiences abnormal latency due to \texttt{checkoutservice-0} being overloaded, the pod \texttt{checkoutservice-0} is the root cause. The service \texttt{CheckoutService} and underlying node \texttt{node-0} may exhibit symptoms, but the actual root cause is at the pod level.

\begin{equation} 
	C^* = \arg\max_{C \in \mathcal{C}} s(C, R)
	\label{eq: fl-def}
\end{equation}

Formally, let $R = \{r_1, r_2, \dots, r_n\}$ denote the set of abnormal requests observed within a diagnostic window, and let $\mathcal{C}$ represent the set of candidate components potentially responsible for these anomalies. The objective of failure localization is to identify the component $C^*$ as defined in Equation~\ref{eq: fl-def}, where $s(C, R)$ is a scoring function that quantifies the likelihood of $C$ being the root cause based on the observed request-level anomalies in $R$.

\subsection{Distributed Tracing} \label{sec:tracing}

Distributed tracing provides end-to-end visibility into request execution across components of a distributed system. It is widely adopted in industry-standard observability stacks (e.g., OpenTelemetry, Jaeger, Zipkin) and has become a foundational data source for latency analysis, root cause diagnosis, and performance optimization.

Each trace records the lifecycle of a single request, capturing all the services and operations involved. A trace is composed of multiple spans, each representing a specific operation such as a database query or RPC invocation. Spans contain structured fields including start/end timestamps, service and operation names, status codes, and a parent\_span identifier, from which parent–child relationships and execution dependencies can be inferred at the trace level.

\begin{figure}[htbp]
	\centering
	\includegraphics[width=1\linewidth]{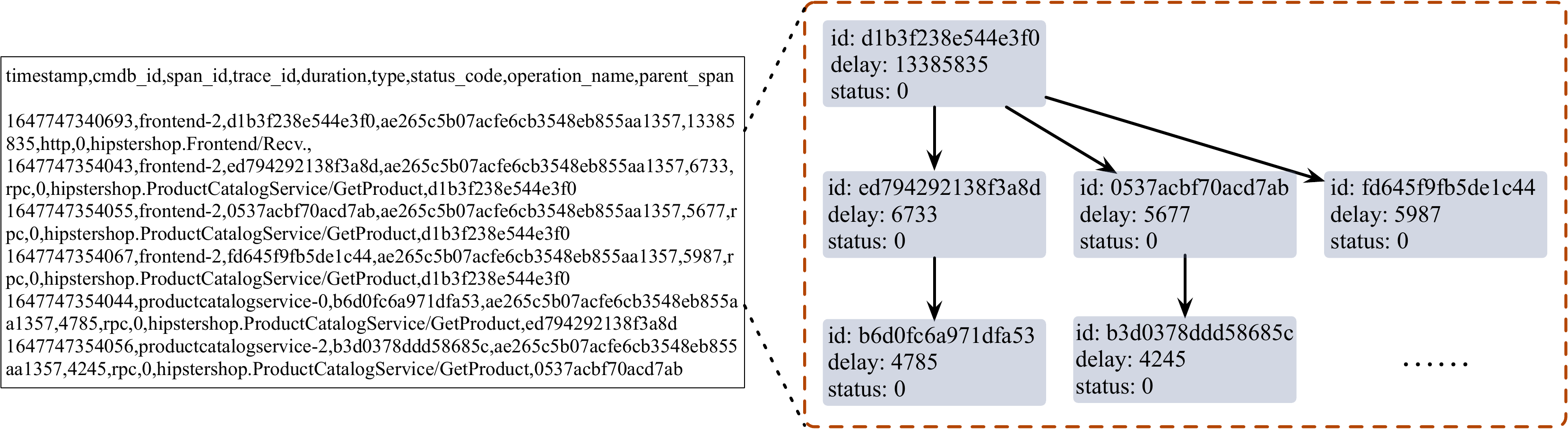}
	\caption{Example of a distributed trace with multiple spans}
	\label{fig: tracing}
\end{figure}

As shown in Figure~\ref{fig: tracing}, a complete trace reconstructs the execution graph of a request. This graph encodes both the control flow and performance dynamics, making it a rich source of context for understanding how and where delays or failures propagate. Entry spans—typically the first span initiated by a frontend or gateway service—are used to determine the health status of the entire request. In our work, we treat a trace as anomalous if (i) the latency of its entry span exceeds 100× the average latency under normal conditions, indicating a performance failure, or (ii) the entry span reports a non-OK status (e.g., non-200/0 codes), indicating a correctness failure. For such anomalous requests—whether caused by performance degradation or correctness violations—localizing the root cause among the many involved services remains a non-trivial but critical task.

\subsection{Reinforcement Fine-Tuning}

Reinforcement Fine-Tuning (RFT) is a paradigm that leverages reinforcement learning (RL) to adapt large language models to complex decision-making tasks in a reward-driven manner~\cite{christiano2017deep, ziegler2019fine}. This approach enables the fine-tuning of models using reinforcement learning techniques to enhance their performance on specialized tasks with minimal training data.

Unlike supervised fine-tuning (SFT), which requires high-quality labeled data and learns via direct instruction, RFT allows models to explore action spaces and learn from reward signals, enabling adaptation in settings where ground truth annotations are scarce or ambiguous. This makes RFT particularly suitable for tasks such as code generation, multi-step reasoning, and tool-use planning.

A number of core algorithms underpin modern RFT strategies. Direct Preference Optimization (DPO)~\cite{rafailov2023direct} simplifies the traditional reward model plus Proximal Policy Optimization (PPO) pipeline by directly optimizing preference-based objectives. PPO~\cite{schulman2017proximal} remains widely used due to its stability in continuous action spaces. Group Relative Policy Optimization (GRPO)~\cite{shao2024deepseekmath} extends PPO by introducing group-based policy updates, allowing for more robust modeling of action diversity.

Next, we briefly introduce the principles of GRPO, which we later extend in our work. GRPO operates over groups of completions, capturing the relative preference structure within each group.

Formally, let $\mathcal{G} = \{(a_1, \dots, a_k)\}$ represent a group of $k$ sampled completions under the same prompt $x$, and let $\pi_\theta$ denote the policy parameterized by $\theta$. GRPO seeks to maximize the group-based expected log-likelihood weighted by soft preferences, as shown in Equation~\ref{eq: grpo1}.

\begin{equation}
	\mathcal{L}_{\text{GRPO}} = \mathbb{E}_{\mathcal{G} \sim \pi_{\theta}} \left[ \sum_{i=1}^{k} w_i \log \pi_{\theta}(a_i|x) \right]
	\label{eq: grpo1}
\end{equation}

Among them, the weights $w_i$ encode the relative preference (or ranking) among completions within the group. These weights are typically computed using a softmax function over reward scores $R(a_i)$, as shown in Equation~\ref{eq: grpo2}, where $\beta$ is a temperature parameter that controls the sharpness of preference.

\begin{equation}
	w_i = \frac{\exp(\beta R(a_i))}{\sum_{j=1}^{k} \exp(\beta R(a_j))}
	\label{eq: grpo2}
\end{equation}

In contrast to PPO’s per-sample KL divergence constraint, GRPO applies a group-wise trust region constraint to regulate policy updates, as described in Equation~\ref{eq: grpo3}.

\begin{equation}
	\text{KL}(\pi_\theta^{\text{group}} \parallel \pi_{\text{old}}^{\text{group}}) \leq \delta
	\label{eq: grpo3}
\end{equation}

This formulation enables more stable and expressive updates by leveraging the relative quality of multiple completions simultaneously, rather than relying on isolated samples. As a result, GRPO is particularly well-suited for tasks involving structured reasoning, where maintaining diversity and avoiding premature convergence to suboptimal outputs is critical.

Recent work has already extended these foundational algorithms to address domain-specific challenges. Token-level DPO (TDPO)~\cite{zeng2024token} introduces token-wise feedback granularity to better align model generation with human intent. Offset-DPO (ODPO)~\cite{amini2024direct} improves training efficiency by assigning non-uniform weights to preference pairs. Completion Pruning PPO (CPPO)~\cite{lin2025cppo} accelerates reasoning tasks by selectively pruning low-quality completions during optimization. Hybrid GRPO~\cite{sane2025hybrid} fuses empirical sampling with value-function stability. Adaptive GRPO (AGPO)~\cite{li2025adaptive} mitigates variance and overthinking issues by modifying advantage estimation and applying length-sensitive reward shaping.

Despite these advances, RFT has not been widely explored in LLM-based observability tasks such as failure localization, where reasoning across complex traces and dynamic dependencies is essential. This motivates our investigation into leveraging RFT to enhance the reasoning ability of lightweight LLMs in practical system diagnosis settings.

\subsection{Key Capabilities for LLM-based Failure Localization}\label{sec:motivation}

\begin{figure}[htbp] 
	\centering 
	\subfigure[Lines Count]{
		 \begin{minipage}{0.48\linewidth} 
		 	\centering 
		 	\includegraphics[width=\textwidth]{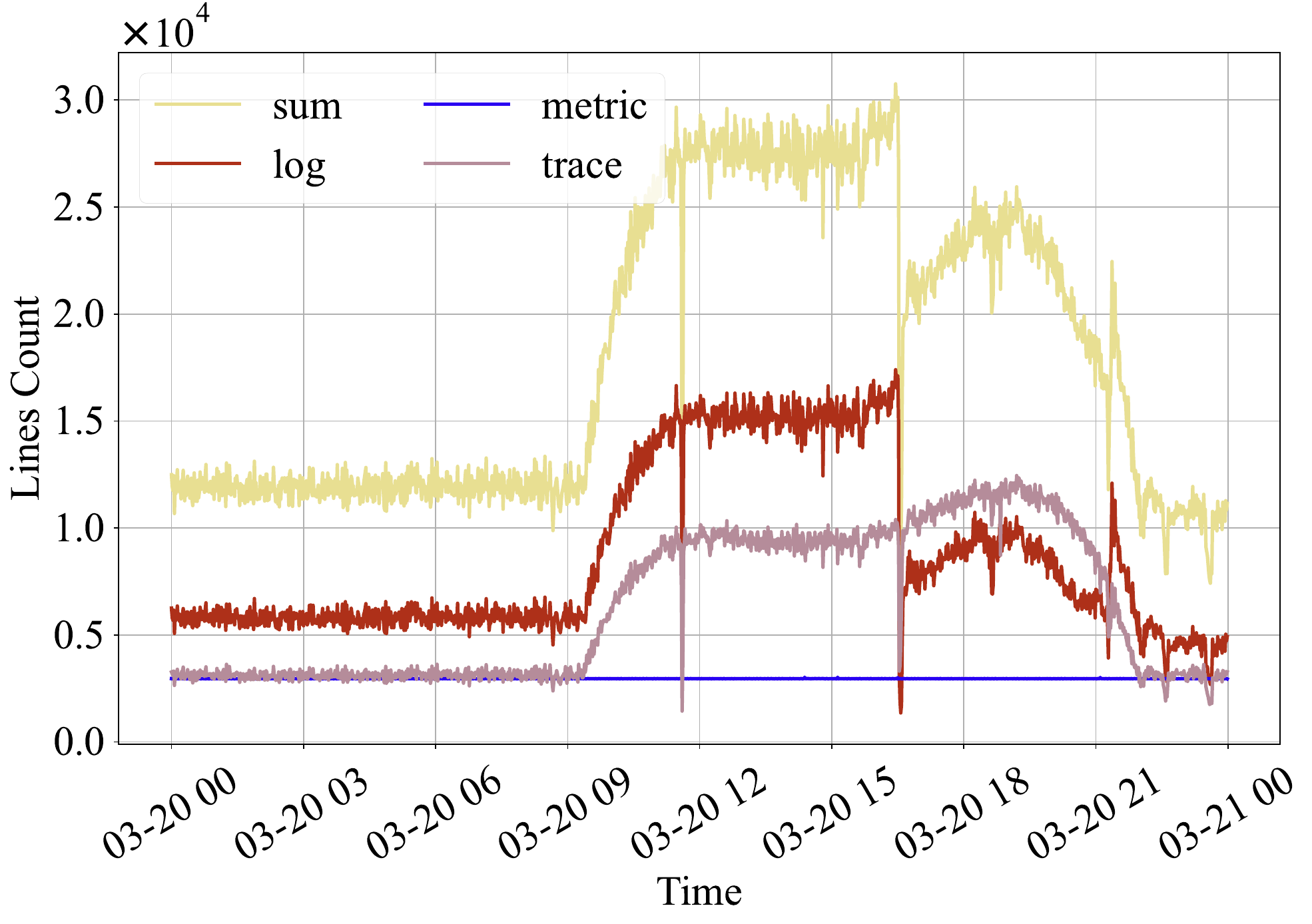} 
		 	\label{fig: dataset-lines} 
	 	\end{minipage} } 
 	\subfigure[Characters Count]{ \begin{minipage}{0.48\linewidth} 
 			\centering 
 			\includegraphics[width=\textwidth]{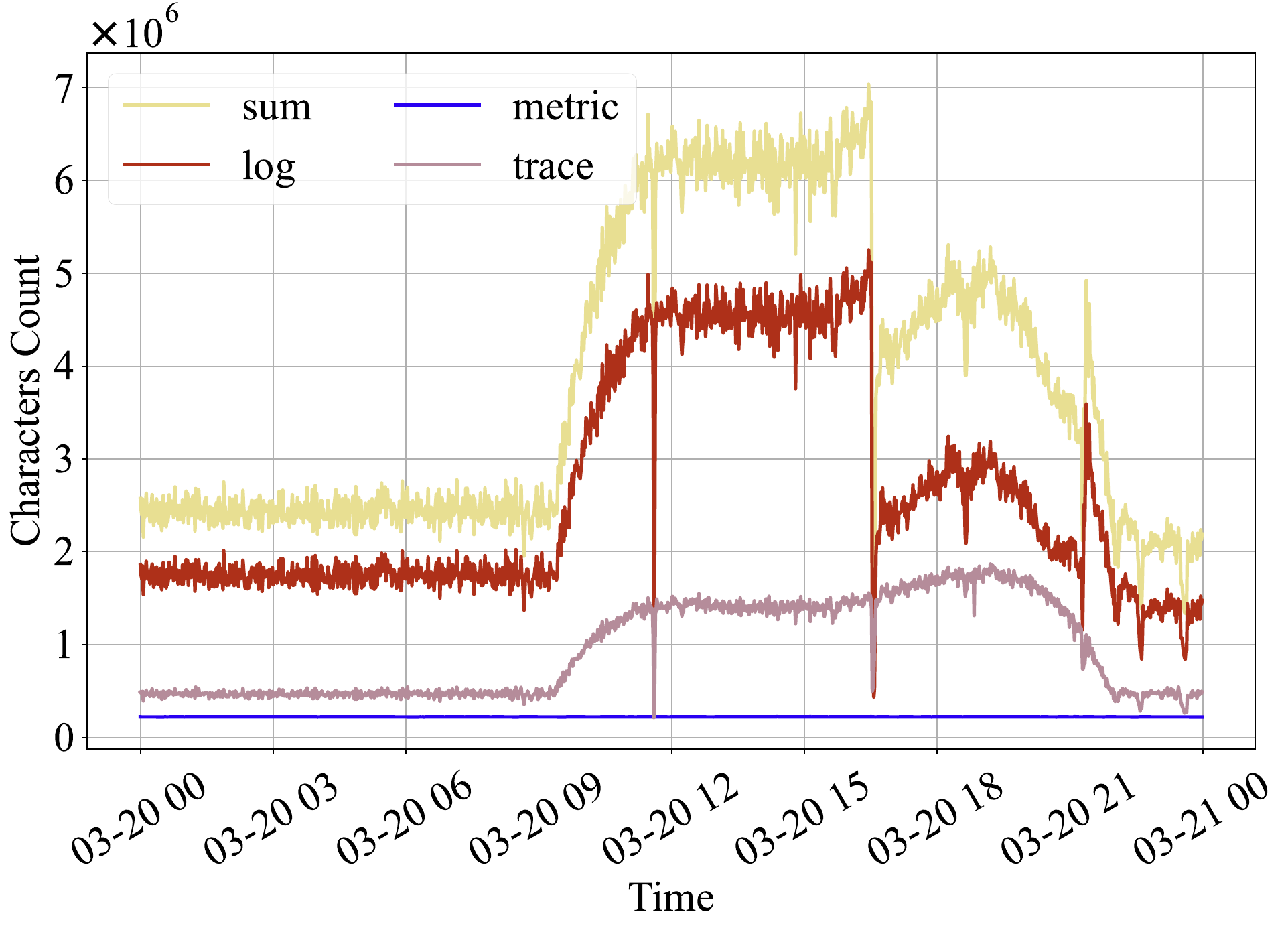}
 			\label{fig: dataset-length} 
 		\end{minipage} } 
 	\caption{Data Source Quantity Within 1-Minute Window} 
 	\label{fig: dataset-analysis} 
\end{figure}

Large-scale distributed systems generate massive volumes of heterogeneous telemetry data—metrics, traces, and logs—especially during failure episodes. As illustrated in Figure~\ref{fig: dataset-analysis}, even a single minute of system activity may produce over 18{,}000 telemetry lines and more than 3.8 million characters. Such scale far exceeds the context limits of modern LLMs, making direct ingestion of raw observations infeasible.

To understand how existing LLM-based failure localization systems cope with this challenge. We gathered relevant papers by searching for work related to LLMs and system-level failure diagnosis, and by following references from closely related studies—including recent surveys on LLM-based automated program repair~\cite{yang2025survey}. We closely examined representative methods, as illustrated in Table~\ref{tab:paper-tools}. These systems span microservice failure diagnosis, production incident analysis, and multi-modality RCA pipelines. While differing in design, they share a common pattern: each augments the LLM with domain-specific tools that preprocess, filter, or reconstruct relevant context prior to reasoning. For example, mABC~\cite{zhang2024mabc}  integrates a dependency query engine and metric explorer to surface suspicious services; RCAgent~\cite{wang2024rcagent} invokes log and trace analysis tools to construct structured evidence; Flow-of-Action~\cite{pei2025flow} coordinates multiple tools under an SOP-style workflow; COCA~\cite{li2025coca} extracts code snippets and execution paths; TAMO~\cite{wang2025tamo} aligns multi-modal observations; and KnowledgeMind~\cite{ren2025multi} pairs metric/log/trace tools with Monte Carlo Tree Search for guided exploration.

\begin{table}[h]
	\centering
	\caption{Summary of representative LLM-based failure localization methods and selected key tools. The table does not include all tools used in each paper, but highlights those that are particularly important.}
	\label{tab:paper-tools}
	\begin{tabular}{cp{10cm}<{\raggedright\arraybackslash}}
		\toprule
		\textbf{Paper} & \multicolumn{1}{c}{\textbf{Main Tools Used}} \\
		\midrule
		mABC~\cite{zhang2024mabc} & 
		\textit{Dependency Query Tool}: Identifies service dependencies to narrow the search space. \newline
		\textit{Metric Tool}: Retrieves key metrics for suspicious services. \\
		\midrule
		
		RCAgent~\cite{wang2024rcagent} &
		\textit{Log Analysis Tool}: Segments long logs and performs RAG-based extraction of evidence for reliable reasoning. \\
		\midrule
		
		Flow-of-Action~\cite{pei2025flow} &
		\textit{Multimodal Data Collection Tool}: Gathers and textualizes metrics, traces, and logs to highlight abnormal signals.\newline
		\textit{SOP Flow Tools}: Structures tool invocation through SOP-guided diagnostic workflows. \\
		\midrule
		
		COCA~\cite{li2025coca} &
		\textit{Code Snippet Extractor}: Locates relevant code regions by log messages.\newline
		\textit{Execution Path Reconstructor}: Rebuilds execution paths leading to the failure to supply actionable context. \\
		\midrule
		
		TAMO~\cite{wang2025tamo} &
		\textit{Multi-modality Alignment Tool}: Aligns diverse telemetry into a unified, time-consistent representation.\newline
		\textit{Failure Localization Tool}: Models causal propagation patterns to pinpoint the fault source.\newline
		\textit{Fault Type Classifier}: Detects failure patterns using a causal graph. \\
		\midrule
		
		KnowledgeMind~\cite{ren2025multi} &
		\textit{Metric Tool}: Detects anomalies in time-series metrics.\newline
		\textit{Log Tool}: Filters and extracts salient log details.\newline
		\textit{Trace Tool}: Examines service-call latencies and communications. \\
		\bottomrule
	\end{tabular}
\end{table}

These observations reveal that the core challenge is not merely {\it having} tools, but enabling the LLM to use them effectively. Across the representative works, two capabilities consistently emerge as fundamental to successful diagnosis:

\paragraph{(1) Tool Invocation Capability.}
The LLM must determine \emph{which} tool to call and \emph{how} to parameterize it in order to distill high-relevance context from massive telemetry streams. Existing systems rely on a variety of such tools—including dependency analyzers, log summarizers, metric explorers, trace analyzers, and multi-modality preprocessors (Table~\ref{tab:paper-tools}). The LLM therefore plays an active role in orchestrating data reduction by selecting and chaining appropriate tools.

\paragraph{(2) Tool Reasoning Chain Capability.}
Failure localization rarely succeeds in a single step; evidence must be gathered iteratively. Modern systems thus couple LLM reasoning with multi-hop tool interactions. Examples include ReAct-style iterative querying (mABC~\cite{zhang2024mabc} , Flow-of-Action~\cite{pei2025flow}), self-consistency aggregation (RCAgent~\cite{wang2024rcagent}), evidence reconstruction pipelines (COCA~\cite{li2025coca}), cross-tool coordination (TAMO~\cite{wang2025tamo}), and search-based planning (KnowledgeMind~\cite{ren2025multi}). In all cases, the LLM must refine hypotheses, resolve contradictions, and issue new tool calls based on previously retrieved information.

\begin{center}
	\begin{tcolorbox}[colback=gray!10, colframe=black, width=\linewidth,
		arc=1mm, auto outer arc, boxrule=0.5pt,
		top=2pt, bottom=2pt, left=2pt, right=2pt]
		\textbf{Summary.} The fundamental difficulty in LLM-based failure localization is not only the sheer volume of telemetry data, but the need for the LLM to orchestrate tools in a multi-step reasoning process. Existing representative methods all embody two essential capabilities: (i) invoking the right tools to distill relevant context, and (ii) forming multi-hop reasoning chains that progressively narrow down the root cause.
	\end{tcolorbox}
\end{center}

\section{Empirical Study}

In this section, we empirically examine whether lightweight LLMs possess the capabilities required for failure localization (as discussed in Section~\ref{sec:motivation}), and whether these capabilities can be enhanced through GRPO-based training.  Our investigation is guided by the following research questions:

\begin{itemize}
	\item \textbf{RQ1:} Do state-of-the-art lightweight LLMs naturally possess the required capabilities?
	\item \textbf{RQ2:} Can existing GRPO algorithms enhance these capabilities and improve failure localization performance?
\end{itemize}

\subsection{Capability of SOTA Lightweight LLMs}

Lightweight LLMs are appealing for failure localization because they are inexpensive and can be deployed efficiently in both edge and cloud environments. However, prior work has consistently shown that such models exhibit fundamental limitations in reasoning and planning~\cite{wu2025inference, zhou2025lightplanner, feng2025efficient}. While most lightweight models---including very small ones such as LLaMA-3.2-3B---can readily acquire tool-invocation syntax through simple function-call fine-tuning~\cite{yuan2025easytool, shen2024llm, yifeilu2025codetool}, syntactic correctness alone does not guarantee effective failure localization. In practice, \textbf{success depends on whether the model can integrate heterogeneous tool outputs into a coherent, multi-step reasoning process}.

To quantify the impact of reasoning capability, we compare lightweight LLMs against a stronger model with demonstrably superior reasoning skills. We reproduced two representative failure-localization methods, mABC and RCAgent, both implemented using LLaMA-3.2-3B as the backbone, and evaluated them on the AIOPS 2022 dataset (subdataset 2022-03-20-cloudbed3). These two methods were chosen because they represent the dominant paradigms in LLM-based system diagnosis: mABC relies on structured, recursive multi-agent pipelines, whereas RCAgent operates through autonomous action--observation loops. To provide an upper bound on achievable performance with stronger reasoning, we additionally ran the same pipelines using Qwen-2.5-Plus.

\begin{table}[htbp]
	\setlength{\tabcolsep}{5pt}
	\centering
	\caption{Failure Localization Results (LLaMA3.2-3B vs.\ Qwen-2.5-Plus)}
	\label{tab: sota-methods}
	\begin{tabular}{c|c|cccccc}
		\toprule
		\multicolumn{2}{c|}{Method} & $Recall@1$ & $Recall@2$ & $Recall@3$ & $Recall@5$ & $Recall@10$ & $MRR$ \\
		\midrule
		\multirow{2}{*}{LLaMA-3.2-3B} & RCAgent & 11.01 & 15.94 & 17.97 & 18.84 & 19.42 & 14.43 \\
		& mABC & 2.65 & 22.26 & 31.17 & 34.54 & 37.79 & 16.83 \\
		\midrule
		\multirow{2}{*}{Qwen-2.5-Plus} & RCAgent & 22.10 & 25.80 & 28.40 & 30.25 & 31.10 & 23.95 \\
		& mABC & 34.19 & 39.87 & 42.13 & 44.51 & 46.78 & 38.46 \\
		\bottomrule
	\end{tabular}
\end{table}

As shown in Table~\ref{tab: sota-methods}, the results highlight several key findings. First, although lightweight models can reliably produce syntactically correct tool calls, they struggle to aggregate and interpret evidence from multiple heterogeneous tools. This limitation is most evident in the performance of mABC with LLaMA-3.2-3B, which attains only 2.65\% Recall@1 despite successfully executing many tool calls. Second, substituting the lightweight backbone with a model possessing stronger reasoning abilities yields substantial performance gains while keeping the tool pipelines unchanged. For example, the Recall@1 of mABC increases from 2.65\% to 34.19\% (+31.54 \%) when LLaMA-3.2-3B is replaced with Qwen-2.5-Plus; RCAgent shows a similar improvement from 11.01\% to 22.10\% (+11.09 \%). These results provide direct quantitative evidence that reasoning capability plays a central role in determining the effectiveness of failure localization. Third, methods with more complex or recursive tool workflows benefit disproportionately from stronger reasoning. mABC exhibits the largest performance jump, reflecting its heavy reliance on multi-hop reasoning patterns that lightweight models fail to execute reliably.

Together, these findings demonstrate that the primary bottleneck in lightweight LLM-based failure localization is not tool usage itself, but the ability to integrate tool outputs into a coherent reasoning chain. Improving lightweight models' reasoning abilities remains essential for advancing practical LLM-driven system diagnosis.

To understand the shortcomings of lightweight models, we conducted a systematic qualitative error analysis on the AIOps 2022 dataset. Specifically, from the full set of 835 failure episodes (including both successful and unsuccessful predictions), we randomly sampled 50 cases in which both lightweight LLMs (LLaMA-3.2-3B) equipped with mABC and RCAgent failed to correctly identify the root cause in their top-1 predictions. For each episode, we examined the full sequence of tool invocations, intermediate outputs, and final predictions. The analysis was performed independently by two researchers with experience in failure localization and LLM-based reasoning, followed by a consensus discussion to resolve discrepancies.

\begin{itemize}
	\item \textbf{Type-1: Lack of termination awareness.}  
	Models frequently fail to recognize when sufficient evidence for localization has been collected. This can result in tool call sequences that are either prematurely truncated—missing critical evidence—or excessively prolonged, introducing noise and confusion into the reasoning context. For example, in diagnosing a failure with span ID \texttt{b711142c1d8f7fd9}, mABC identified an anomaly in the downstream span ID \texttt{6a214f194590145c} and prematurely concluded the reasoning process, treating the corresponding RecommendationService as the root cause and proceeding with further analysis. In reality, the true root cause was located three layers deeper, in the ProductCatalogService.
	
	\item \textbf{Type-2: Parameter extraction failure.}  
	When the output of previous tools becomes lengthy or semantically complex, lightweight models struggle to correctly extract the necessary fields for the next tool call. For instance, the service name or metric type inferred from a diagnostic trace may be incorrectly formatted, omitted, or mismatched. For example, RCAgent is built upon log analysis and includes adaptations for handling lengthy log data. However, the processed log blocks can still contain hundreds of lines, making it difficult for lightweight models to extract potential root causes for the next reasoning step. As a result, critical information is often missed.
	
	\item \textbf{Type-3: Missed root cause despite available evidence.}  
	Even in relatively simple cases where the first two issues do not occur—that is, the relevant data for the true root cause is successfully retrieved during the tool invocation process—lightweight models sometimes still fail to correctly identify it. For example, during mABC's reasoning for a failure with span ID \texttt{b562d708ca6f300f}, all potential root cause services were retrieved, including Frontend, CheckoutService, and EmailService, yet the model incorrectly ranked the true root cause outside the top-1 prediction.
\end{itemize}

A closer comparison between the two methods further reveals why their failure patterns differ. For the 50 mispredicted cases, mABC exhibits 3 Type-1 failures, 5 Type-2 failures, and 42 Type-3 failures. This distribution aligns with its design characteristics: mABC relies primarily on a voting-based mechanism with a relatively strict and stable reasoning path and a more tolerant parameter-extraction rule (i.e., minor extraction failures do not affect the final decision). As a result, it is less likely to suffer from Type-1 or Type-2 issues, but heavily impacted by Type-3 failures.

By contrast, RCAgent shows 19 Type-1 failures, 7 Type-2 failures, and 24 Type-3 failures. Because RCAgent must autonomously plan its reasoning trajectory and decide when to terminate, and because long multi-turn interactions may accumulate excessive context, it is more vulnerable to Type-1 and Type-2 errors. While its retry mechanism mitigates parameter-extraction issues to some degree, its flexible reasoning flow makes it more likely to deviate when encountering ambiguous or noisy environment feedback.

Notably, both methods exhibit a substantial number of Type-3 failures, which is expected: locating the correct root cause in long, noisy, and highly redundant contexts remains fundamentally challenging for lightweight models.

These observations suggest that the core bottleneck is not tool-use syntax but insufficient semantic understanding and reasoning planning. Lightweight LLMs, due to their limited capacity, are especially susceptible to cascading failures in multi-step reasoning. Unlike larger models that can generalize stable reasoning templates across tasks and remain coherent even under imperfect tool outputs, smaller models often collapse once intermediate steps diverge from their expected distribution.

\begin{center}
	\begin{tcolorbox}[colback=gray!10,
		colframe=black,
		width=\linewidth,
		arc=1mm, auto outer arc,
		boxrule=0.5pt,
		top=2pt,
		bottom=2pt,
		left=2pt,
		right=2pt
		]
		\textbf{Summary.} Although lightweight LLMs can acquire tool invocation abilities through simple tuning, their effectiveness in failure localization remains limited. The key missing link is their inability to construct coherent, goal-driven reasoning chains under noisy and multi-modal data conditions.
	\end{tcolorbox}
\end{center}

\subsection{Direct Application of GRPO Algorithm}

To further enhance lightweight LLMs with the ability to perform deep and strategic failure localization through tool usage, we experimented with applying the Group Relative Policy Optimization (GRPO) algorithm—originally proposed in the DeepSeekMath work~\cite{shao2024deepseekmath} and subsequently used in the training of DeepSeek-R1~\cite{guo2025deepseek}—to train LLaMA3.2-3B and LLaMA3-8B models. The key motivation is to leverage reward-based fine-tuning to refine the tool invocation strategy in complex reasoning chains. Given that the core metric for failure localization is Mean Reciprocal Rank (MRR), we set the reward function directly to MRR and implemented GRPO using the OpenRLHF framework~\cite{hu2024openrlhf}. We selected LLaMA-3.2-3B and LLaMA-3.1-8B as the GRPO training backbones for three pragmatic reasons. First, they are representative lightweight open-source models, covering the small ($\approx$3B) and medium ($\approx$8B) parameter regimes most relevant for real-world, cost-sensitive deployment. Second, the LLaMA family benefits from a large user community and a mature tooling ecosystem for fine-tuning and RL workflows, which simplifies implementation and reproducibility of GRPO experiments. Third, and importantly, LLaMA’s pretraining and tokenization are strongly English-centric, which better matches the language characteristics of our observability signals (service names, logs, and trace annotations). In our preliminary internal comparisons, LLaMA backbones produced more stable baseline tool-invocation behavior than some other lightweight families, so applying GRPO to these models allowed us to more clearly isolate and measure the gains due to reinforcement fine-tuning. Taken together, these considerations made LLaMA-3.2-3B and LLaMA-3.1-8B a practical and informative choice for evaluating GRPO in the lightweight regime.

\paragraph{Training Setup.}
Following the standard GRPO pipeline, we adopted a two-phase training regime:
\begin{itemize}
	\item \textbf{Phase 1: Supervised Fine-Tuning (SFT).} We used a small-scale synthetic dataset composed of multi-turn diagnostic dialogues generated by Claude-3.5-Sonnet. This step imparts initial tool invocation capabilities and coarse-grained decision heuristics.
	\item \textbf{Phase 2: GRPO-Based RL Fine-Tuning.} The models were then further optimized with GRPO, using MRR as the reward signal. GRPO refines the model by evaluating the relative performance of the group of candidate policies, encouraging outputs that improve their group-relative rank. Instead of relying on traditional reinforcement learning techniques like value or critic networks, GRPO directly optimizes the policy by comparing the relative performance of multiple candidates, thus enabling more robust exploration and policy improvement.
\end{itemize}

\begin{figure}[htbp]
	\centering
	\subfigure[Reward]{
		\begin{minipage}{0.48\linewidth}
			\centering   
			\includegraphics[width=\textwidth]{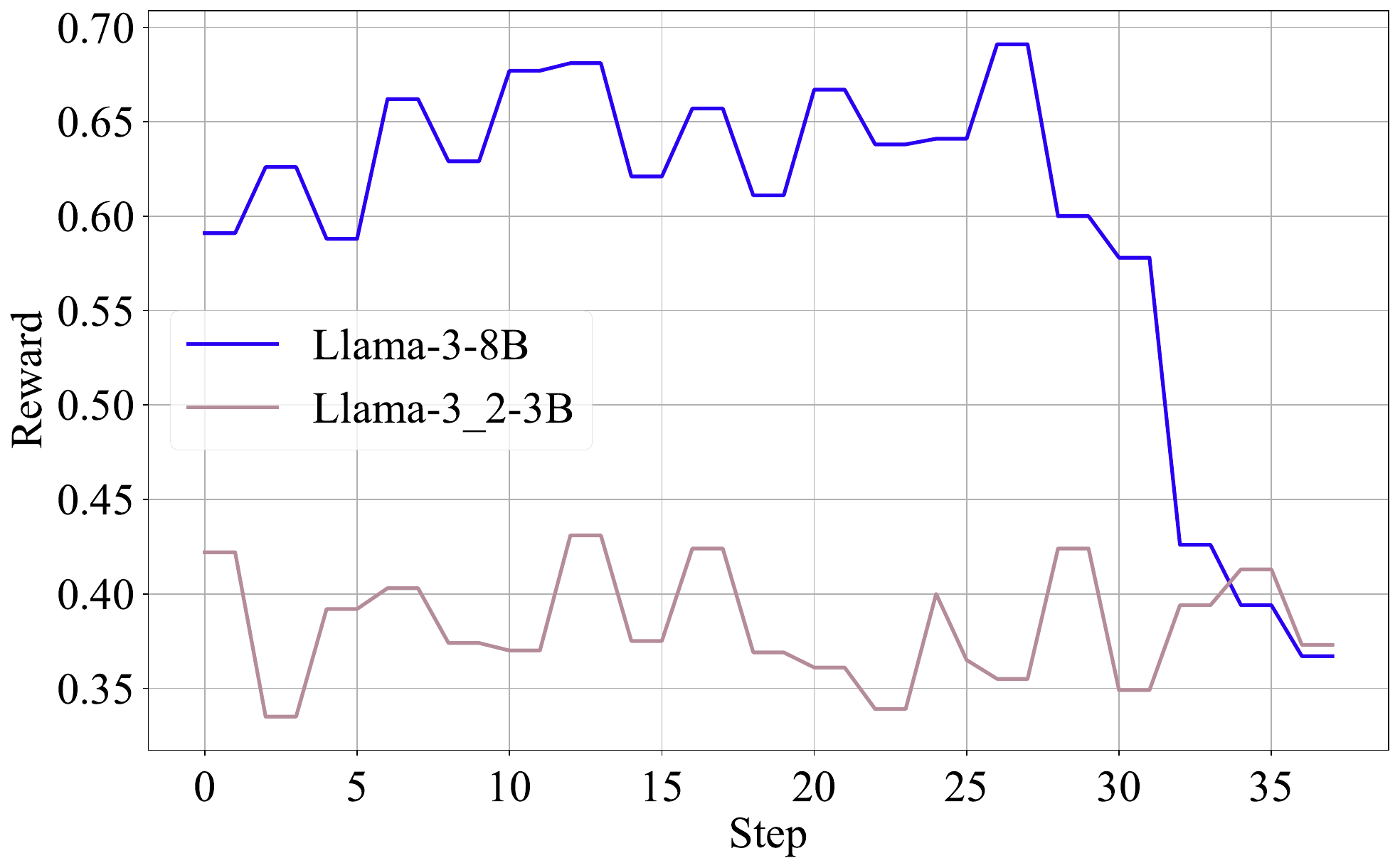}
			\label{fig:reward}
		\end{minipage}
	}
	\subfigure[Response Length]{
		\begin{minipage}{0.48\linewidth}
			\centering
			\includegraphics[width=\textwidth]{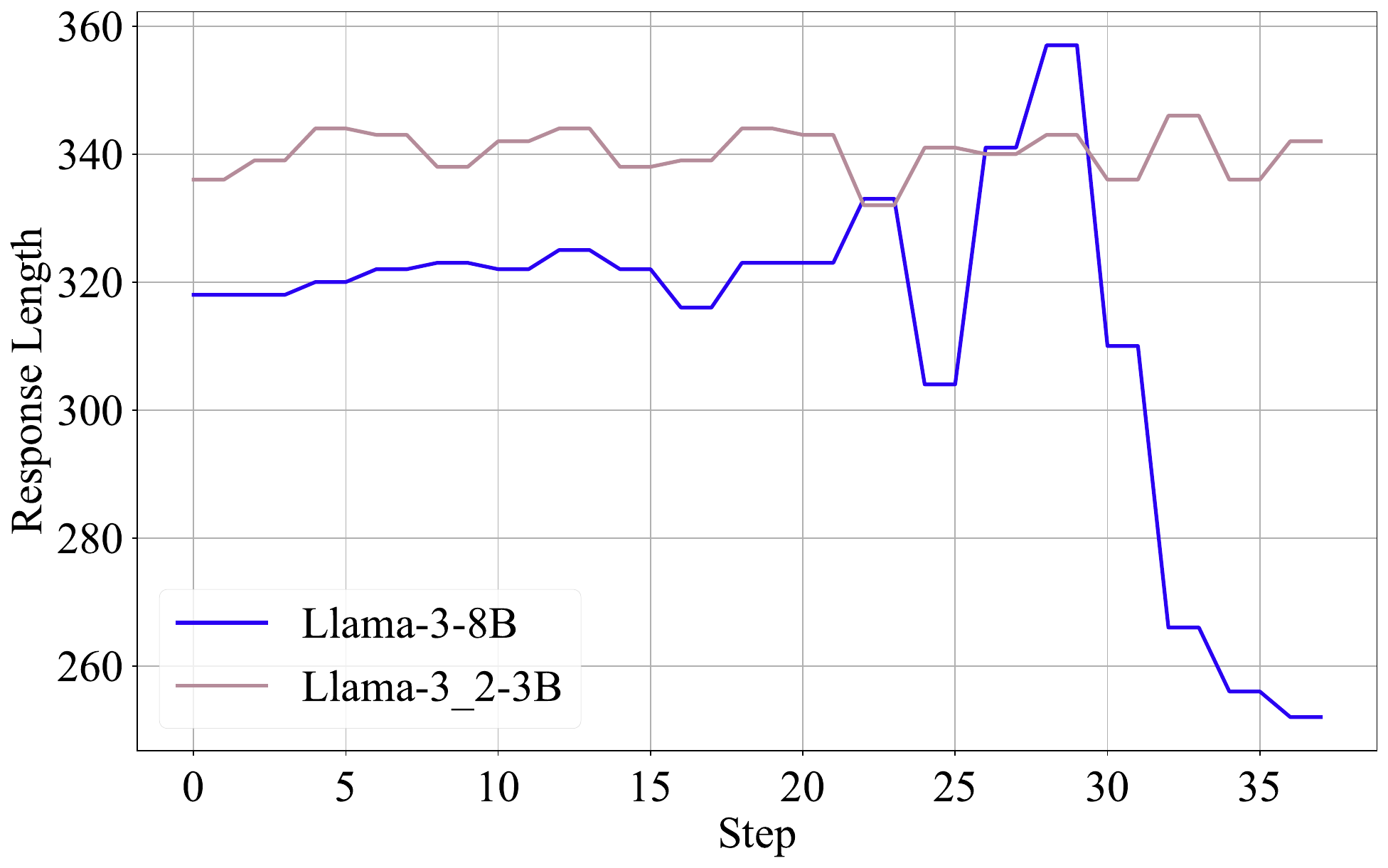}
			\label{fig:response-length}
		\end{minipage}
	}
	\caption{Training Effects of Direct GRPO Application}
	\label{fig:direct-application}
\end{figure}

\paragraph{Training Dynamics.}
As illustrated in Figure~\ref{fig:direct-application}, the training curves show limited or even negative improvement in reward and response diversity:
\begin{itemize}
	\item For \textbf{LLaMA3-8B}, we observed a steady \textit{decline} in both the average reward and the response length. This implies that the model learned to avoid complex tool chains and instead favored shorter, low-risk responses—a clear symptom of \textit{reward hacking}.
	\item The \textbf{LLaMA3.2-3B} model maintained a nearly constant reward (~0.38) throughout training, failing to discover improved strategies or leverage exploration to escape local optima.
\end{itemize}

\paragraph{Reward Signal Limitation.}
A core issue lies in the nature of the MRR reward signal itself. MRR is inherently \textbf{non-smooth and discontinuous}:
\begin{itemize}
	\item A correct answer ranked first yields a score of 1, but this drops to 0.5 if ranked second, and to 0.33 for third, etc.
	\item This sharp drop-off produces a highly imbalanced reward landscape, where most sampled outputs yield near-zero rewards, making it difficult for policy updates to gain signal and direction.
\end{itemize}

\paragraph{Qualitative Observations.}
Inspection of generated outputs post-training revealed several patterns:
\begin{itemize}
	\item Models tend to collapse to single-step predictions, avoiding exploration of longer reasoning paths.
	\item When multi-step outputs were generated, their structure was often shallow or repetitive, failing to refine hypotheses based on intermediate tool feedback.
	\item Tool parameters were often valid syntactically but lacked task-relevance, reflecting a disconnection between search strategy and task-specific semantics.
\end{itemize}

\begin{center}
	\begin{tcolorbox}[colback=gray!10,
		colframe=black,
		width=\linewidth,
		arc=1mm, auto outer arc,
		boxrule=0.5pt,
		top=2pt,
		bottom=2pt,
		left=2pt,
		right=2pt
		]
		\textbf{Summary.} The direct application of the GRPO algorithm fails to effectively steer lightweight LLMs toward more optimal failure localization strategies. The coarse and imbalanced MRR-based reward signal hinders effective exploration, while the models’ low capacity limits their ability to recover from early-stage reward sparsity.
	\end{tcolorbox}
\end{center}

In summary, our empirical study reveals that in order to equip lightweight LLMs with failure localization capabilities, the following challenges must be addressed:

\begin{itemize}
	\item \textbf{High Dependence on Standardized Tool Invocation.} In failure localization, directly feeding all raw data into the LLM rarely yields the correct root cause. Instead, LLM-based methods depend heavily on data querying tools. Since the localization process often requires multiple tool invocations—each based on the results of previous queries—any deviation in the standardized format of these invocations can mislead the model’s judgment. This is in stark contrast to the knowledge QA or mathematical reasoning tasks where GRPO is typically applied. 
	\item \textbf{Demand for Exploration under Standardized Tool Usage.} Although standard supervised fine-tuning (SFT) combined with tool learning can teach LLMs to use tools correctly, such constraints on output format can greatly limit the model’s exploratory capabilities. In contrast, failure localization requires the LLM to explore diverse ways of invoking tools—while still maintaining standardization—to discover the correct pathway under different scenarios. 
	\item \textbf{Integrated Reward Mechanism for Reasoning Process and Final Output.} Once the tool-based inference of the root cause is complete, the final evaluation cannot rely solely on semantic correctness, as in typical knowledge QA tasks. Instead, it necessitates a multi-dimensional scoring approach that combines a rating of the final root cause, the quality of the reasoning path, and penalties for hallucinations. 
\end{itemize}

\section{ThinkFL}

The observations in empirical study underscore the need for a dedicated failure localization paradigm that moves beyond static tool-usage templates. It must support flexible, multi-turn tool reasoning, provide feedback-grounded rewards, and enable lightweight models to improve iteratively over time.

\begin{figure*}[htbp] 
	\centering \includegraphics[width=1\linewidth]{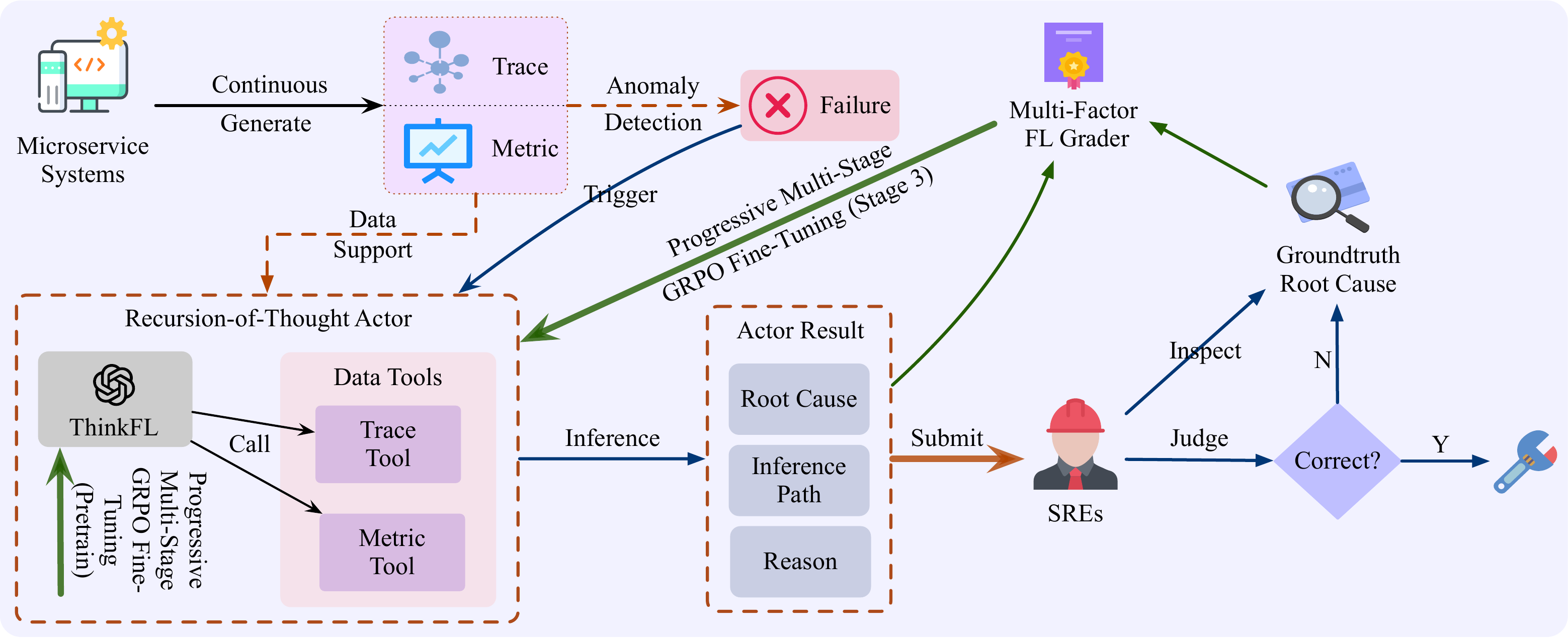} \caption{Pipeline of ThinkFL} 
	\label{fig: pipeline} 
\end{figure*}

Therefore, in this section, we introduce \textbf{ThinkFL}, a self-refining failure localization model for microservice systems using reinforcement fine-tuning. ThinkFL is designed to bridge the gap between static tool-use policies and dynamic, feedback-driven optimization by combining structured tool invocation with progressive, reward-sensitive adaptation. Figure~\ref{fig: pipeline} illustrates the overall workflow of ThinkFL.

As microservice systems operate continuously, they generate large volumes of trace and metrics data. An external anomaly detection module—which is not part of ThinkFL itself—first processes this data to determine whether a request exhibits abnormal behavior. In our implementation, this step follows the heuristic described earlier in Section~\ref{sec:tracing}, though in practice it may be replaced with any more advanced anomaly detection technique.

Once an anomalous trace is identified, the workflow transitions to failure localization, where ThinkFL begins its reasoning from the entry span of the trace, which anchors the diagnostic episode. For example, an entry span may look like: \texttt{timestamp=1647747340693, service=frontend-2, traceID=d1b3f238, spanID=ae265c5b, duration=13385835, protocol=http, status=0, operation=hipstershop.Frontend/Recv.} Each field provides essential context: the timestamp of the span, the service instance processing the request, the trace ID and span ID identifying the request and its sub-operation, the duration (µs), the protocol used, the status code, and the operation name. From this input, ThinkFL invokes a \textbf{Recursion-of-Thought (RoT) Actor}, which interacts with multiple data tools, including trace analysis and metric evaluation tools. The RoT Actor ultimately produces an actor result, consisting of a ranked list of potential root causes, corresponding explanations, and a complete inference path that is easily interpretable by humans.

The generated actor result is then submitted to expert SREs for validation. If the inferred result is correct, the SRE proceeds with remediation. If it is incorrect, the SRE uses the provided inference path and the list of ranked candidate services as additional context when identifying the true root cause. A \textbf{Multi-Factor Failure Localization (FL) Grader} is then applied, incorporating both the generated actor result and the verified ground-truth root cause to evaluate ThinkFL’s performance. Finally, ThinkFL undergoes further refinement through a \textbf{Progressive Multi-Stage GRPO Fine-Tuning} approach, enhancing its ability to localize failures more accurately over time.

\subsection{Recursion-of-Thought Actor}

\begin{figure}[htbp] 
	\centering 
	\includegraphics[width=0.6\linewidth]{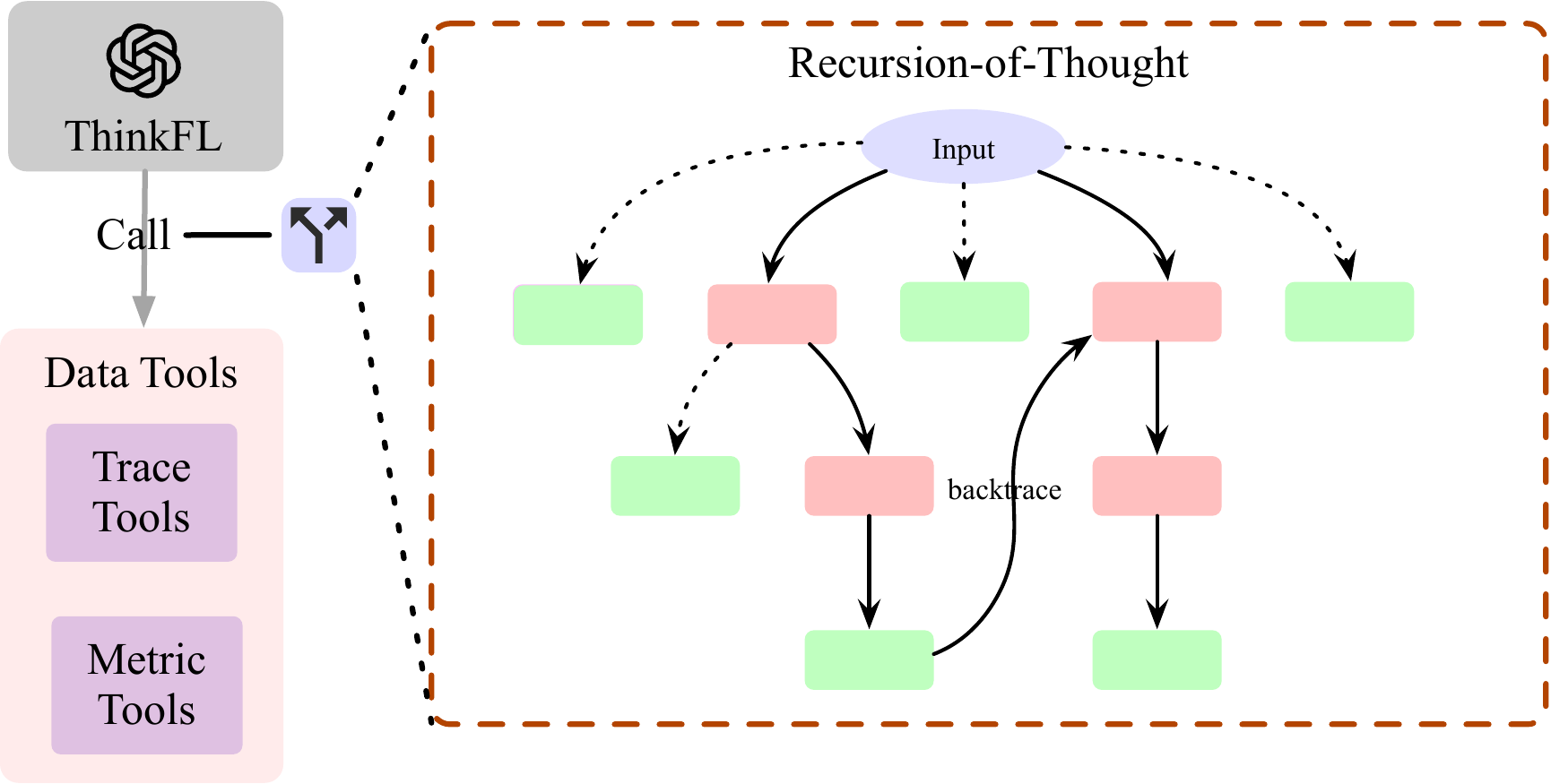} 
	\caption{Workflow of Recursion-of-Thought Actor} 
	\label{fig: RoT-Actor} 
\end{figure}

Within the overall workflow, the core reasoning engine is the Recursion-of-Thought Actor, as illustrated in Figure~\ref{fig: RoT-Actor}. This actor is built upon a pre-trained large language model—\textit{ThinkFL}—which continues to evolve through online refinement and task-specific feedback. Its primary role is to perform recursive, multi-step reasoning by selectively invoking a diverse set of data tools and integrating their results into the evolving inference state.

Unlike static prompting pipelines, the Recursion-of-Thought Actor operates in a dynamic, context-aware loop: at each step, it evaluates the current evidence, decides which data tool to apply next, and updates its internal reasoning chain accordingly. Through this iterative process, the actor progressively gathers and synthesizes relevant information, ultimately converging on a ranked set of likely root causes. This design enables flexible adaptation to a wide range of failure patterns and system configurations.

To support this reasoning loop, a collection of \textit{data tools} serves as the interface between the actor and the underlying system telemetry. These tools abstract the complexity of large-scale observability data and provide structured, filtered insights that facilitate efficient and focused analysis.

\subsubsection{Data Tools}

Data tools are modular components that provide structured access to specific types of system data. Given a query input, each tool retrieves and preprocesses the relevant information, making it readily consumable by the reasoning agent. In our current implementation, we focus on two primary tools: the Trace Tool and the Metrics Tool. However, the framework is extensible, allowing additional tools—such as logs, configuration diffs, or service topology graphs—to be integrated as needed.

\textbf{Trace Tool.} Distributed trace data captures the end-to-end flow of requests across microservices, making it a foundational resource for root cause localization. In such traces, a \textit{span} represents a single unit of work, typically corresponding to the execution of a service operation or a method within a microservice. Each span records key information such as its start and end timestamps, the service handling the request, the operation name, and status codes. Spans are organized hierarchically, where a parent span can trigger multiple \textit{child spans}—these \textit{span-level dependencies} encode the causal relationships between service calls, effectively forming a tree or directed acyclic graph (DAG) of the request flow across microservices.

To illustrate, consider a request processed by a frontend service that generates span ID \texttt{b711142c}. This request triggers the CheckoutService, which generates a direct child span with ID \texttt{c4829a7f}, and the CheckoutService further invokes the PaymentService, producing another child span \texttt{d9357b2a}. Here, \texttt{b711142c} is the parent span, \texttt{c4829a7f} is its direct child, and \texttt{d9357b2a} is a child of the CheckoutService span. Understanding these span-level dependencies is critical: the true root cause of a failure may reside in a downstream service, and reasoning over the span hierarchy allows the localization agent to trace how errors propagate through the microservice system.

However, the sheer volume of trace data—especially in systems with high request throughput—makes it infeasible to process all spans. To address this, we introduce a Trace Tool that performs targeted retrieval of relevant spans based on span-level dependencies:

\begin{equation}
	T(s) = \left\{ \langle t, s', svc, op, d, \sigma \rangle \; \middle| \; s' \in \mathcal{C}(s) \right\}
	\label{eq: trace-agent}
\end{equation}

Formally, for a given span identifier \(s\), the Trace Tool returns the set of its direct child spans \(\mathcal{C}(s)\), along with associated metadata such as timestamp \(t\), child span identifier \(s'\), service name \(svc\), operation name \(op\), duration \(d\), and status code \(\sigma\). This selective retrieval allows the reasoning agent to incrementally explore the trace graph, capturing causal relationships between microservices while staying within the context limits of the LLM.

\textbf{Metrics Tool.} Metrics reflect the performance and health status of various system components, such as services, pods, or nodes. However, in large-scale environments, the volume of metrics can be overwhelming, and most remain stable even during system faults. To reduce noise and focus on actionable signals, we design a Metrics Tool that selectively surfaces only those metrics exhibiting statistically significant deviations.

\begin{equation}
	|m(t) - \mu_m| > n \times \sigma_m
	\label{eq: n-sigma}
\end{equation}

Specifically, given a timestamp \(t_0\) and a target component \(C\), let \(\mathcal{M}(C)\) be the set of associated metrics. For each metric \(m \in \mathcal{M}(C)\), we calculate its historical mean \(\mu_m\) and standard deviation \(\sigma_m\). The Metrics Tool applies an \(n\)-sigma test over the window \([t_0 - \delta, t_0 + \delta]\) to detect abnormal fluctuations:

\begin{equation}
	Q(t_0, \delta, C) = 
	\left\{ 
	m(t) \;\middle|\; 
	\begin{array}{c}
		m \in \mathcal{M}(C), \\
		t \in [t_0 - \delta, t_0 + \delta], \\
		|m(t) - \mu_m| > n \times \sigma_m
	\end{array} 
	\right\}
	\label{eq: metric-agent}
\end{equation}

Crucially, the Metrics Tool is not a standalone anomaly detector. It is invoked in the context of a suspected failure episode, where the Trace Tool has already identified candidate components along the span hierarchy. The Metrics Tool then queries only those components and only within the failure’s time window, surfacing deviations that are most relevant to the ongoing episode. In this way, the tool \textit{links failures to metrics} by conditioning the retrieval on (i) the failure trigger time and (ii) the components under investigation.

This design ensures that the reasoning process is not overwhelmed by irrelevant data, while still leveraging quantitative signals that reinforce or contradict hypotheses generated during trace exploration. Together, the Trace and Metrics Tools provide complementary perspectives—traces identify \emph{where} failures propagate, while metrics explain \emph{why} they occur. Combined, they form the foundation of the Recursion-of-Thought Actor’s reasoning capability, enabling it to navigate vast diagnostic spaces in a scalable and interpretable manner.

\subsubsection{Recursion-of-Thought}

While traditional Chain-of-Thought (CoT) approaches have shown effectiveness in structured reasoning tasks, they encounter substantial challenges when applied to failure localization in complex distributed systems~\cite{ren2025multi}. In these environments, the diagnostic space is large, noisy, and hierarchically structured—rendering linear reasoning brittle and prone to premature convergence on misleading signals. In contrast, experienced site reliability engineers (SREs) rarely follow a rigid path. Instead, they iteratively formulate and revise hypotheses, consult various data sources, and backtrack as needed.

Inspired by this human-like diagnostic behavior, we propose the Recursion-of-Thought (RoT) framework—a dynamic and self-correcting reasoning paradigm tailored for root cause inference. Unlike CoT, which constructs a static sequence of reasoning steps, RoT performs recursive reasoning driven by evolving context. As illustrated in Algorithm~\ref{alg: recursion-of-thought}, where the LLM iteratively selects one of the tools—Trace, Metrics, or Print—based on the evolving reasoning path. Each selected tool contributes new evidence to the inference path $R$, which is then used to guide the next step. The algorithm proceeds recursively with a bounded depth, and if no conclusive root cause is printed within the limit, it triggers a fallback mechanism to generate the final output based on the accumulated reasoning history.

\begin{algorithm}[htbp]
	\caption{Recursion-of-Thought Algorithm}
	\label{alg: recursion-of-thought}
	\begin{algorithmic}[1]
		\Require Entry trace \(T\) for a high-duration request.
		\Ensure Formatted potential root cause(s) \(R_f\).
		
		\State Initialize inference path \(R \gets \{T\}\)
		\State Initialize allowed actions \(\mathcal{A} \gets \{ \text{Trace}, \text{Metrics}, \text{Print} \}\)
		\State \(\textit{printed} \gets \text{False}\)
		\State Set maximum recursion depth \(d \gets D_{\max}\)
		
		\While{\(d > 0\)}
		\State Generate reasoning instruction \(I \gets f(R, \mathcal{A})\)
		\State \(action, params \gets \text{Decide}(I)\)
		
		\If{\(action = \text{Trace}\)}
		\State \(D \gets \text{TraceTool}(params)\)
		\State \(R \gets R \cup \{D\}\)
		\ElsIf{\(action = \text{Metrics}\)}
		\State \(M \gets \text{MetricsTool}(params)\)
		\State \(R \gets R \cup \{M\}\)
		\ElsIf{\(action = \text{Print}\)}
		\State \(R_f \gets \text{PrintRootCauses}(R)\)
		\State \(\textit{printed} \gets \text{True}\)
		\State \Return \(R_f\)
		\EndIf
		
		\State \(d \gets d - 1\)
		\EndWhile
		
		\If{\(\textit{printed} = \text{False}\)}
		\State \(R_f \gets \text{PrintRootCauses}(f(R, \{\text{Print}\}))\)
		\State \Return \(R_f\)
		\EndIf
	\end{algorithmic}
\end{algorithm}

\paragraph{Tool Invocation Strategy.}

Rather than relying on predefined heuristics or action scoring, the Recursion-of-Thought Actor adopts a fully LLM-driven strategy for tool selection. As illustrated in Figure~\ref{fig: tool-invocation-prompt}, we provide a feasible prompt that demonstrates this process (the actual prompt used in practice includes additional parameters required by the data tools as well as intermediate reasoning steps, making it more complex). At each reasoning step, the actor leverages the contextual knowledge embedded in the language model to decide which tool to invoke—or whether to proceed directly with inference.

\begin{figure}[htbp]
	\centering
	\begin{tcolorbox}[colback=gray!10,
		colframe=black,
		width=\linewidth,
		arc=1mm, auto outer arc,
		boxrule=0.5pt,
		top=2pt, bottom=2pt, left=2pt, right=2pt
		]
		Please continue to identify the root cause service. 
		You may explore deeper by using the \texttt{search\_traces} tool or combine with 
		\texttt{search\_fluctuating\_metrics}. 
		If you have determined the root cause, call the \texttt{print\_results} function.
		
		Available tools:\\
			\textbf{print\_results}: report candidate root causes (node/service/pod) with reasoning. \\
			\textbf{search\_traces}: retrieve child spans of a given \texttt{span\_id}. \\
			\textbf{search\_fluctuating\_metrics}: retrieve anomalous metrics around a given \texttt{service\_name} and timestamp.
	\end{tcolorbox}
	\caption{Illustrative prompt for tool invocation.}
	\label{fig: tool-invocation-prompt}
\end{figure}

This strategy capitalizes on the LLM’s ability to interpret partial observations, maintain reasoning continuity, and adapt its behavior based on newly acquired evidence. Given the current inference path and relevant context, the model formulates its next move through natural language reasoning, without relying on external prioritization logic.

Such a flexible and data-driven decision mechanism ensures generalizability across diverse scenarios, and enables seamless integration of new tools or reasoning patterns without modifying the core workflow.

\paragraph{Interpretability of the Reasoning Path.}
Beyond enhancing inference performance, the RoT framework offers high interpretability by maintaining an explicit and structured reasoning path \(R\). Each recursive step—including the generated instruction, the invoked tool, and its output—is appended to this path, forming a transparent trace of the decision-making process.

This interpretability supports expert validation, facilitates post-mortem diagnosis, and enables knowledge distillation for future learning cycles. As a result, the framework not only provides accurate inferences but also fosters trust and reproducibility in system-level debugging workflows.

\subsection{Multi-Factor Failure Localization Grader}

\begin{figure}[htbp] 
	\centering \includegraphics[width=0.6\linewidth]{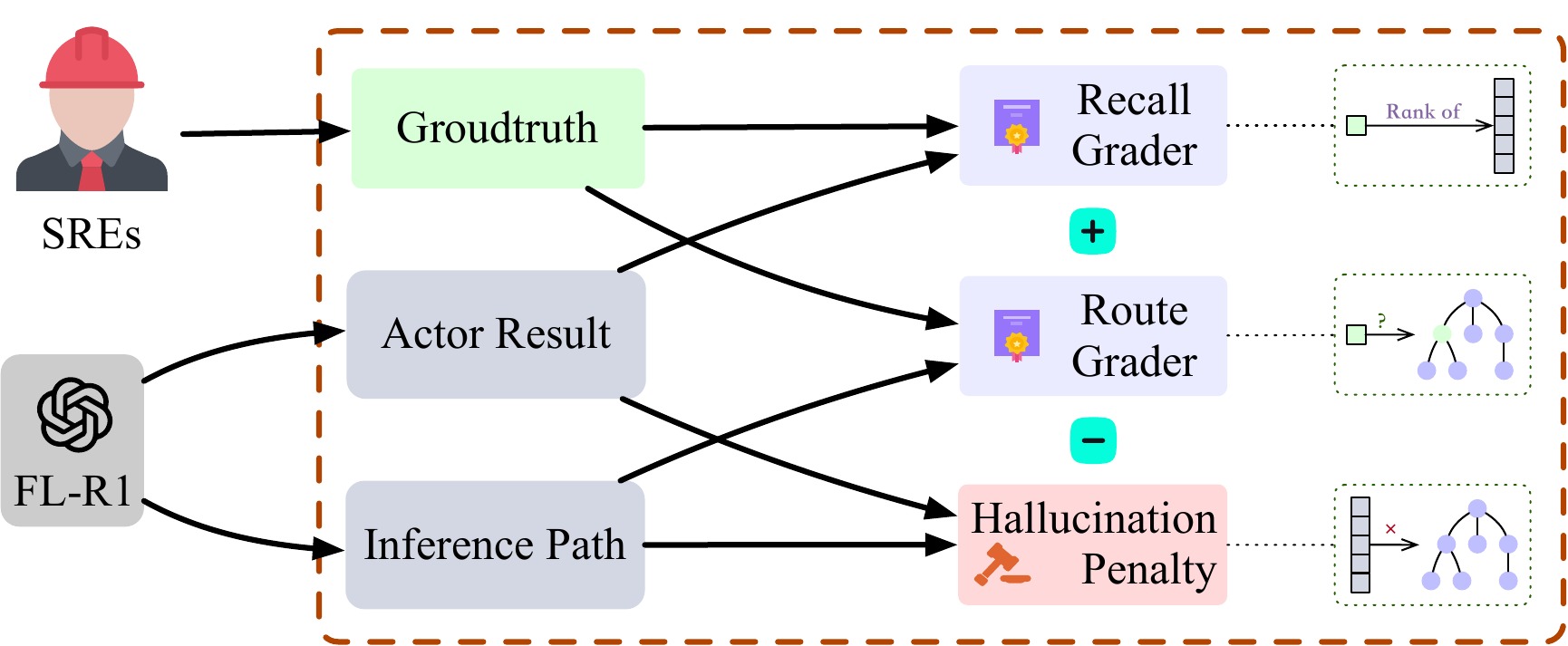} 
	\caption{Structure of Multi-Factor Failure Localization (FL) Grader} 
	\label{fig: Multi-Factor-FL-Grader} 
\end{figure}

During the online execution of the entire workflow through the recursion-of-thought algorithm, the core component is the multi-factor failure localization grader, as illustrated in Figure~\ref{fig: Multi-Factor-FL-Grader}. This grader evaluates the model's performance from multiple dimensions: (1) Result Correctness via the Recall Grader, (2) Reasoning Path Validity via the Route Grader, and (3) Answer Reliability through a Hallucination Penalty, which discourages the generation of unsupported or fabricated content. The overall score is computed as a weighted combination of these components, formulated in Equation~\ref{eq: grader}, where $R$ denotes the recall score, $P$ represents the route matching score, and $H$ is the hallucination penalty. The weights $\alpha$, $\beta$, and $\gamma$ control the relative influence of each component.

\begin{equation} 
	\label{eq: grader} 
	S = \alpha \cdot R + \beta \cdot P - \gamma \cdot H
\end{equation}

\subsubsection{Recall Grader}

The recall grader is the most critical component for directly evaluating the model’s final output. While the final failure localization performance is measured using metrics like MRR, this type of reward is often too sparse to effectively guide the training of RoT models. As illustrated in our empirical study, discrete rewards such as MRR provide limited feedback signals during training.

\begin{equation} 
	\label{eq: recall-grader} 
	R = \begin{cases} 
		1 - \frac{r}{r_{max}}, & r \leq r_{\text{max}} \\ 
		\frac{1}{r_{max}}, & r \textgreater r_{\text{max}}
	\end{cases}
\end{equation}

To address this, we adopt a linear rank-based scoring scheme, as formulated in Equation~\ref{eq: recall-grader}, where $r$ is the rank of the correct root cause in the model’s predicted list. We define a maximum effective rank $r_{max}$, beyond which all predictions receive a fixed low score. For ranks within the valid range $[0, r_{max}]$, each higher position in the ranking receives a proportionally lower score, with rank 0 corresponding to a full score of 1.

\subsubsection{Route Grader}

The Route Grader is designed to evaluate the correctness and appropriateness of the reasoning path generated during failure localization. It serves two main purposes. First, it measures the position of the correct root cause within the reasoning path: a later appearance (closer to the end) suggests a well-refined inference. However, we introduce a tolerance threshold to account for slightly extended reasoning paths, which may include normal service components that help confirm the diagnostic conclusion. Second, if the correct root cause is not present in the path, the grader instead assigns a score based on the path length, under the assumption that a longer reasoning process may still yield useful evidence.

\begin{equation} 
	\label{eq: route-grader} 
	P = \begin{cases} 
		\min\left\{\frac{r}{L - \mu},\ 1\right\}, & \text{if } r \in R \\ 
		\min\left\{\frac{L}{D_{\text{max}}},\ 1\right\}, & \text{if } r \notin R
	\end{cases}
\end{equation}

To capture these two perspectives, the route score \(P\) is defined as a combination of the position score \(P_{\text{pos}}\) and the length score \(P_{\text{len}}\). The position score reflects how close the correct root cause appears to the end of the reasoning path. If it occurs within the final \(\mu\) steps of a reasoning path of length \(L\), a full score of 1 is awarded; otherwise, the score decays linearly. The length score rewards longer reasoning paths, up to a maximum threshold \(D_{\text{max}}\), encouraging more comprehensive analysis when the root cause is not identified. In summary, the route grader can be calculated as Equation~\ref{eq: route-grader}.

\subsubsection{Hallucination Penalty}

During training, we observed that lightweight LLMs occasionally generate hallucinated information when exposed to unfamiliar data. Specifically, two major types of hallucinations frequently occur in the final root cause outputs:  (1) predicting root causes that do not appear in the generated reasoning path, and  (2) redundantly outputting the same root cause multiple times.

\begin{equation}
	\label{eq: hallucination-penalty}
	H = \lambda_1 \cdot \frac{N_{\text{inv}}}{N_{\text{total}}} + \lambda_2 \cdot \frac{N_{\text{dup}}}{N_{\text{total}}}
\end{equation}

To penalize these behaviors and encourage the model to produce faithful and diverse outputs, we introduce a hallucination penalty, as defined in Equation~\ref{eq: hallucination-penalty}. Here, \(N_{\text{inv}}\) denotes the number of predicted root causes not found in the reasoning path, \(N_{\text{dup}}\) is the number of duplicated root causes, \(N_{\text{total}}\) is the total number of predicted root causes, and \(\lambda_1\) and \(\lambda_2\) are weighting coefficients used to balance the penalty for each type of hallucination.

\subsection{Progressive Multi-Stage GRPO Fine-Tuning}

\begin{figure*}[htbp] 
	\centering \includegraphics[width=1\linewidth]{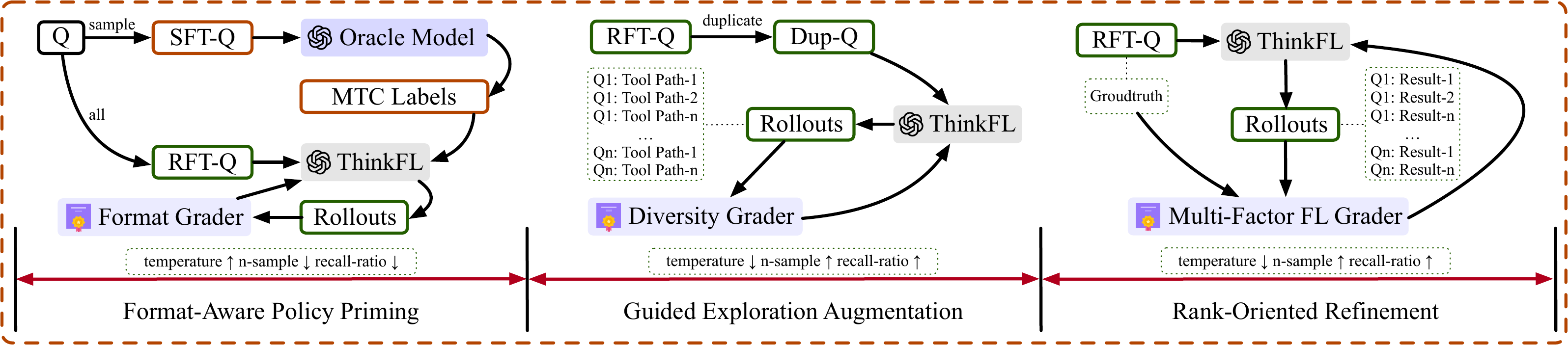} \caption{Workflow of Progressive Multi-Stage GRPO Fine-Tuning} 
	\label{fig: Progressive-GRPO} 
\end{figure*}

Although the proposed multi-factor failure localization grader offers effective guidance for improving root cause identification via Recursion-of-Thought (RoT), the lightweight LLM backbone still exhibits two prominent limitations: (1) suboptimal external tool invocation behavior, and (2) insufficient diversity in reasoning trajectories. To mitigate these issues, we introduce a progressive multi-stage GRPO fine-tuning algorithm, designed to incrementally enhance the model's capabilities.

As depicted in Figure~\ref{fig: Progressive-GRPO}, the proposed framework consists of three carefully orchestrated stages:
\begin{enumerate}
	\item \textbf{Format-Aware Policy Priming}, which equips the model with accurate and structured tool usage abilities;
	\item \textbf{Guided Exploration Augmentation}, which encourages diverse and informative reasoning behaviors;
	\item \textbf{Rank-Oriented Refinement}, which aligns the model's outputs with high-quality, grader-validated reasoning trajectories.
\end{enumerate}

\subsubsection{Format-Aware Policy Priming}

This stage aims to establish a robust foundation for tool-augmented reasoning. It operates in two phases: \textbf{Supervised Fine-Tuning (SFT)} and \textbf{Reinforced Fine-Tuning (RFT)}.

\textbf{SFT Phase.} We begin by sampling a small, diverse subset of questions from the raw dataset. Broadly speaking, the raw dataset consists of accumulated, high-quality historical failure localization cases, including the questions, reasoning process, and answers. In this work, the raw dataset specifically refers to the AIOPS 2022 dataset, from which we select only 100 questions for the SFT phase. Using an oracle model (e.g., Claude or DeepSeek-R1), we generate multi-turn conversations (MTCs) that follow the Recursion-of-Thought (RoT) protocol. These MTCs include both intermediate reasoning steps and the final root cause predictions, serving as high-quality supervision to teach the lightweight LLM (ThinkFL) how to invoke tools correctly and structure its reasoning.

Importantly, we keep the SFT scale deliberately small to preserve the model’s exploratory capacity and avoid premature overfitting to rigid patterns.

\textbf{RFT Phase.} Building on the SFT initialization, the model enters a reinforced learning stage in which it explores diverse reasoning paths under higher sampling temperature. For each generated trajectory, we apply a hybrid reward function comprising:
\begin{itemize}
	\item A recall grader, which measures correctness of the final prediction;
	\item A format grader, which evaluates structural adherence to the expected schema.
\end{itemize}

The format grader plays a critical role in ensuring output consistency and tool usability. It verifies schema compliance across multiple levels, such as correct inclusion of \texttt{name} and \texttt{arguments} fields, valid JSON-like argument structures, and well-formed root cause elements containing essential attributes like \texttt{node}, \texttt{service}, and \texttt{pod}. Malformed outputs are down-weighted to enforce format discipline.

\begin{algorithm}[htbp]
	\caption{Diversity Grader Evaluation}
	\label{alg: diversity-grader}
	\begin{algorithmic}[1]
		\Require Question $q$, Reasoning path $p$, Path cache $\mathcal{P}_q$
		\State $p' \gets$ \textsc{Deduplicate}($p$)
		\If{$p'$ not in $\mathcal{P}_q$}
		\State $\mathcal{P}_q \gets \mathcal{P}_q \cup \{p'\}$
		\If{\textsc{SolvesProblem}($q$, $p'$)}
		\State \Return Score $= A$ \Comment{New + Correct}
		\Else
		\State \Return Score $= B$ \Comment{New but Incorrect}
		\EndIf
		\Else
		\State \Return Score $= C$ \Comment{Repetitive}
		\EndIf
	\end{algorithmic}
\end{algorithm}

\subsubsection{Guided Exploration Augmentation}

While the priming stage focuses on correctness and format, this stage emphasizes \textbf{reasoning diversity}, which aims to enhance the model’s ability to explore diverse failure localization paths while maintaining correct tool invocation. To achieve this, we introduce data augmentation through controlled duplication of input questions. By presenting the same question multiple times during training, we encourage the model to generate varied reasoning paths and corresponding root causes across different training steps. Each generated output is evaluated using a combined grading mechanism consisting of a diversity grader and a recall grader, with a slightly higher weight assigned to the recall grader to discourage the model from exploiting the reward signal by generating overly noisy or irrelevant reasoning paths.

The diversity grader operates as described in Algorithm~\ref{alg: diversity-grader}. For each question, we maintain a dynamic cache of previously generated reasoning paths. After a new reasoning path is produced, redundant tool invocations—those with identical tools and parameters—are filtered out. The path is then compared against the cache: if the path is entirely novel, a high reward \(A\) is given; if the path solves the question with only slight variation, a moderate reward \(B\) is assigned; if the path is already present in the cache, a lower reward \(C\) is used, where \(A \gg B > C\).

\subsubsection{Rank-Oriented Refinement}

After the previous two stages, the model has developed a strong capability in failure localization. The goal of this final stage—Rank-Oriented Refinement—is to further consolidate the model’s reasoning and output quality. For each question, the model generates multiple rollouts, and these outputs are subsequently evaluated using the multi-factor failure localization grader.

As aforementioned, this grader comprehensively assesses each rollout based on multiple dimensions, such as accuracy of root cause identification, reasoning trace completeness, and format correctness. By ranking the outputs according to these scores and reinforcing the higher-quality responses, the model is guided to converge towards more precise and reliable failure localization strategies.

\section{Evaluation}

To evaluate the effectiveness and robustness of ThinkFL, we design a series of experiments guided by the following research questions:

\begin{itemize}
	\item \textbf{EV-RQ1:} How does ThinkFL perform in failure localization accuracy compared to baseline LLMs and state-of-the-art failure localization approaches?
	\item \textbf{EV-RQ2:} What is the inference efficiency of ThinkFL in terms of runtime, and how does it compare with baseline LLMs?
	\item \textbf{EV-RQ3:} How does each stage in the ThinkFL training pipeline contribute to the final localization accuracy?
	\item \textbf{EV-RQ4:} What are the key behavioral and policy changes observed during the GRPO training process, and how do they reflect the model’s evolving decision strategies?
	\item \textbf{EV-RQ5:} How effective is the proposed ThinkFL training method when applied to different LLM backbones (e.g., Qwen2.5, Llama3), and what is the trade-off between model size and performance?
	\item \textbf{EV-RQ6:} In what ways does self-refinement in ThinkFL enable more accurate and nuanced failure localization, as illustrated through detailed case studies?
	\item \textbf{EV-RQ7:} How sensitive is ThinkFL’s performance to key hyperparameters?
	\item \textbf{EV-RQ8:} How well does ThinkFL generalize to other microservice systems?
\end{itemize}

\subsection{Experimental Setup}

\subsubsection{Dataset} 

Our experiments use the AIOPS 2022 dataset from a mature microservices-based e-commerce system, spanning six subsets across two days and three cloudbeds; for clarity, we denote them as $\mathbf{A}$, $\mathbf{B}$, $\mathbf{\Gamma}$, $\mathbf{\Delta}$, $\mathbf{E}$, $\mathbf{Z}$. The monitored system has 7 microservices deployed over 44 pods and 6 nodes, with traces ranging from hundreds to thousands of spans.

We define failures at the end-to-end request level. A trace is anomalous if its entry span latency exceeds 100× the normal average or reports an abnormal status (e.g., 5xx). Each failure is annotated with the ground-truth root cause at exactly one level: service (e.g., CheckoutService), pod (e.g., checkoutservice-0), or node (e.g., node-0).

Each case provides (i) the entry span with basic attributes, (ii) the full distributed trace (span DAG), (iii) time-aligned node/service/pod metrics around the failure, and (iv) the labeled root cause. This format supports traversal of span-level dependencies and validation using metrics, facilitating methods like ThinkFL for component-level localization.

\subsubsection{Baseline Models and Approaches}

We compared ThinkFL against five state-of-the-art (SOTA) LLMs and eight representative root cause localization methods. The LLMs—Claude-3.5-Sonnet, Qwen-2.5-Plus, Llama3.1-70B, DeepSeek-R1-Qwen-32B, and Qwen-2.5-Max—were guided to reason under the Recursion-of-Thought (RoT) paradigm to generate final root causes.

The eight SOTA localization approaches can be categorized into two groups: ML/DL-based, and LLM-based methods.

\textbf{ML/DL-based:} CRISP~\cite{zhang2022crisp} represents traces as critical paths and applies a lightweight heuristic to identify root cause instances. TraceContrast~\cite{zhang2024trace} utilizes sequence representations, contrastive sequential pattern mining, and spectrum analysis to localize multi-dimensional root causes. TraceRank~\cite{yu2023tracerank} combines spectrum analysis with a PageRank-based random walk algorithm to pinpoint anomalous services. MicroRank~\cite{yu2021microrank} constructs a trace coverage tree to capture dependencies between requests and service instances, leveraging the PageRank algorithm to score potential root causes. RUN~\cite{lin2024root} employs time series forecasting for neural Granger causal discovery and integrates a personalized PageRank algorithm to efficiently recommend the top-k root causes. Microscope~\cite{lin2018microscope} builds causality graphs and utilizes a depth-first search strategy to detect front-end anomalies.

\textbf{LLM-based:} mABC~\cite{zhang2024mabc} proposes a multi-agent, blockchain-inspired collaboration framework where multiple LLM-based agents follow a structured workflow and collaborate through blockchain-inspired voting mechanisms. RCAgent~\cite{wang2024rcagent} is an LLM-based reasoning and acting framework where a controller agent executes a thought–action–observation loop and dynamically invokes expert agents as tools for domain-specific tasks, with a flexible exit mechanism to report findings.

\subsubsection{Evaluation Metrics}

We use the top-k recall (Recall@k) and mean reciprocal rank (MRR) to evaluate the accuracy of root cause localization following existing works~\cite{zhang2024trace}.

\begin{itemize}
	\item \textbf{Recall@$k$:} Measures the proportion of fault instances for which the ground-truth root cause appears in the top-$k$ positions of the predicted list. Formally, as illustrated by Equation~\ref{eq: recallk}, given a set of fault instances $A$, let $Hit@k_i$ denote whether the root cause for the $i$-th instance appears within the top-$k$ predictions ($Hit@k_i = 1$ if yes, $0$ otherwise). In this paper, we evaluate Recall@1, Recall@2, Recall@3, Recall@5, and Recall@10.
	\begin{equation}
		Recall@k = \frac{1}{|A|} \sum_{i=1}^{|A|} Hit@k_i
		\label{eq: recallk}
	\end{equation}
	\item \textbf{MRR:} is the multiplicative inverse of the rank of the root cause in the result list. If the root cause is not included in the result list, the rank can be regarded as positive infinity. Given a set of fault instances $A$, $Rank_i$ is the $i$ rank of the root cause in the returned list of the $i$th fault instance, MRR is calculated by Equation~\ref{eq: mrr}.
\end{itemize}

\begin{equation}
	MRR = \frac{1}{|A|} \sum_{i=1}^{|A|} \frac{1}{Rank_i}
	\label{eq: mrr}
\end{equation}

\subsubsection{Implementation and Settings}

We implement our algorithm using OpenRLHF v0.6.1.post1~\cite{hu2024openrlhf}. Unless otherwise specified, we set the maximum effective rank to $r_{\text{max}} = 10$, the maximum dialogue depth to $D_{\text{max}} = 10$, and the weighting parameters to $\alpha = 1$, $\beta = 0.2$, and $\gamma = 0.2$. Throughout all experiments, we adopt Llama-3.2-3B as the backbone LLM for fine-tuning. For comparison, all state-of-the-art LLM-based failure localization baselines are executed on Qwen-2.5-Plus.

All experiments are conducted on a CentOS 8 Linux server equipped with 24 Intel(R) Xeon(R) CPUs (2.90GHz), 400GB of RAM, and two NVIDIA A800 GPUs, each with 80GB of memory.

\subsubsection{Evaluation Procedure}

For each research question (EV-RQ1–EV-RQ8), we follow a systematic procedure to ensure fair and reproducible evaluation:

\begin{itemize}
	\item \textbf{Data selection:} For RQs evaluating accuracy and efficiency (EV-RQ1 and EV-RQ2), all six subsets of the AIOPS 2022 dataset are used. For generalization studies (EV-RQ8), the TrainTicket dataset is additionally evaluated. Each subset is treated independently to capture different workload patterns and system behaviors.
	
	\item \textbf{Method execution:} Each baseline and ThinkFL model is executed on the corresponding dataset subset. LLM-based methods are prompted using the Recursion-of-Thought or tool-invocation paradigms as described in Section~3. For trace- and metrics-based methods, the original or reproduced implementations are applied directly to the subset.
	
	\item \textbf{Metric computation:} Recall@k and MRR are computed for each subset-method combination. For multi-step methods (e.g., ThinkFL), final predictions are aggregated per failure case to compute metrics. This ensures that evaluation reflects both accuracy and practical usability.
	
	\item \textbf{Result aggregation:} Metrics are averaged across all subsets for overall comparison, while subset-level results are reported to analyze sensitivity to workload, system configuration, and dataset characteristics. Ablation studies and hyperparameter sensitivity analyses are similarly performed per subset before aggregation.
\end{itemize}

\subsection{Overall Accuracy}

\begin{table}[htbp]
	\setlength{\tabcolsep}{7pt}
	\centering
	\caption{Evaluation Results Compared with SOTA LLMs}
	\label{tab:overall-recall}
	\begin{tabular}{c|c|cccccc}
		\toprule
		\multicolumn{2}{c|}{Model} & $\mathbf{A}$ & $\mathbf{B}$ & $\mathbf{\Gamma}$ & $\mathbf{\Delta}$ & $\mathbf{E}$ & $\mathbf{Z}$ \\
		\midrule
		\multirow{6}{*}{Claude-3.5-Sonnet} 
		& $Recall@1$  & 28.47 & 41.26 & 36.29 & 38.06 & 30.40 & 39.87 \\
		& $Recall@2$  & 59.66 & 57.40 & 69.10 & 49.74 & 54.19 & 62.41 \\
		& $Recall@3$  & 62.37 & 70.85 & 89.22 & 64.86 & 64.76 & 72.66 \\
		& $Recall@5$  & 64.41 & 72.20 & 91.73 & 71.74 & 69.82 & 76.58 \\
		& $Recall@10$  & 68.14 & 73.54 & 92.57 & 79.35 & 78.19 & 80.63 \\
		& $MRR$ & 46.13 & 54.42 & 60.07 & 51.86 & 48.51 & 56.17 \\
		\midrule
		
		\multirow{6}{*}{Qwen-2.5-Plus} 
		& $Recall@1$  & 21.11 & 26.82 & 26.63 & 15.79 & 11.89 & 26.93 \\
		& $Recall@2$  & 30.79 & 26.82 & 49.15 & 20.29 & 34.58 & 32.08 \\
		& $Recall@3$  & 32.18 & 27.27 & 50.24 & 24.58 & 37.00 & 33.63 \\
		& $Recall@5$  & 32.87 & 28.18 & 53.51 & 29.39 & 40.75 & 35.56 \\
		& $Recall@10$  & 33.91 & 29.55 & 58.35 & 35.67 & 45.15 & 39.05 \\
		& MRR & 26.89 & 27.37 & 39.98 & 21.73 & 25.75 & 31.08 \\
		\midrule
		
		\multirow{6}{*}{Llama3.1-70B} 
		& $Recall@1$  & 16.67 & 26.46 & 23.74 & 25.44 & 6.17 & 24.69 \\
		& $Recall@2$  & 30.42 & 26.91 & 45.08 & 28.99 & 27.88 & 29.77 \\
		& $Recall@3$  & 32.92 & 28.25 & 48.44 & 32.85 & 28.95 & 33.45 \\
		& $Recall@5$  & 35.83 & 29.60 & 53.12 & 35.35 & 34.05 & 36.78 \\
		& $Recall@10$  & 38.33 & 31.84 & 59.23 & 39.10 & 41.82 & 43.43 \\
		& $MRR$ & 25.69 & 27.88 & 37.74 & 29.79 & 19.92 & 30.44 \\
		\midrule
		
		\multirow{6}{*}{DeepSeek-R1-Qwen-32B} 
		& $Recall@1$  & 14.34 & 14.81 & 27.30 & 18.81 & 12.66 & 15.77 \\
		& $Recall@2$  & 27.91 & 14.81 & 50.42 & 21.67 & 38.23 & 22.43 \\
		& $Recall@3$  & 27.91 & 15.34 & 51.53 & 23.10 & 39.75 & 23.73 \\
		& $Recall@5$  & 29.46 & 15.87 & 53.76 & 24.05 & 41.77 & 27.21 \\
		& $Recall@10$  & 33.72 & 17.99 & 60.31 & 28.81 & 46.82 & 32.71 \\
		& $MRR$ & 22.19 & 15.56 & 40.98 & 21.76 & 27.33 & 21.36 \\
		\midrule
		
		\multirow{6}{*}{Qwen-2.5-Max} 
		& $Recall@1$  & 3.20 & 16.51 & 7.80 & 5.75 & 2.08 & 10.36 \\
		& $Recall@2$  & 13.52 & 18.87 & 36.76 & 7.51 & 30.09 & 16.87 \\
		& $Recall@3$  & 14.23 & 19.81 & 37.75 & 8.40 & 32.18 & 18.46 \\
		& $Recall@5$  & 15.30 & 20.75 & 39.23 & 10.50 & 33.79 & 20.05 \\
		& $Recall@10$  & 18.86 & 22.17 & 41.21 & 11.27 & 35.18 & 21.65 \\
		& $MRR$ & 9.78 & 18.42 & 23.54 & 7.56 & 17.37 & 14.93 \\
		\midrule
		
		\multirow{6}{*}{ThinkFL-3B (\textit{ours})} 
		& $Recall@1$  & 37.63 & 60.54 & 49.34 & 60.90 & 33.04 & 51.52 \\
		& $Recall@2$  & 65.42 & 71.30 & 81.20 & 75.29 & 55.95 & 74.05 \\
		& $Recall@3$  & 72.54 & 73.54 & 86.71 & 80.60 & 69.38 & 77.47 \\
		& $Recall@5$  & 73.56 & 75.34 & 90.54 & 83.73 & 76.43 & 81.14 \\
		& $Recall@10$  & 74.92 & 75.78 & 91.86 & 86.86 & 80.62 & 84.05 \\
		& $MRR$ & \textbf{54.44} & \textbf{67.13} & \textbf{68.26} & \textbf{71.05} & \textbf{51.48} & \textbf{65.22} \\
		\bottomrule
	\end{tabular}
\end{table}

We first compare ThinkFL against state-of-the-art LLMs. As shown in Table~\ref{tab:overall-recall}, ThinkFL significantly outperforms all baseline models on the failure localization task across all evaluation metrics and datasets. In terms of $Recall@1$, ThinkFL achieves an average improvement of 13.05\% over Claude-3.5-Sonnet, the strongest among the compared baselines, highlighting its superior ability to identify the correct root cause at the top-ranked position. On dataset $\mathbf{B}$, ThinkFL attains 60.54\% in $Recall@1$, a substantial lead over Claude-3.5-Sonnet's 41.26\%, suggesting its stronger discriminative capability in more complex cases. Similarly, for dataset $\mathbf{\Delta}$, ThinkFL improves $Recall@1$ by over 22.84\% compared to Claude-3.5-Sonnet, demonstrating robustness even under challenging diagnostic scenarios.

As the number of retrieved candidates increases, ThinkFL consistently maintains its lead. At $Recall@3$ and $Recall@5$, ThinkFL records average gains of 5.88\% and 5.70\% over Claude-3.5-Sonnet, respectively. These results indicate that even when multiple hypotheses are allowed, ThinkFL provides more relevant candidate causes, reinforcing its utility in real-world scenarios where analysts often consider a short list of likely causes. Moreover, ThinkFL exhibits outstanding performance in Mean Reciprocal Rank (MRR), with an average improvement of 10.08\% over Claude-3.5-Sonnet. The MRR advantage indicates that ThinkFL not only ranks the correct answer more frequently at the top, but also returns it earlier in the list more consistently than any other model.

In addition to outperforming Claude-3.5-Sonnet, ThinkFL demonstrates significant superiority over other strong baselines such as Qwen-2.5-Plus and Llama3.1-70B. On dataset $\mathbf{\Gamma}$, for instance, ThinkFL reaches 86.71\% in $Recall@3$, compared to 50.24\% and 48.44\% for Qwen-2.5-Plus and Llama3.1-70B, respectively—illustrating its notable edge in identifying root causes in high-noise settings. On dataset $\mathbf{E}$, which requires more intricate and multi-hop reasoning to accurately identify the root cause, ThinkFL achieves an MRR of 51.48\%, outperforming Qwen-2.5-Plus and Llama3.1-70B by 25.73\% and 31.56\%, respectively. These substantial margins underscore ThinkFL’s superior capacity for deep reasoning and causal inference, particularly in scenarios where surface-level correlations are insufficient and a nuanced understanding of system behavior is essential for correct localization.

We further compare ThinkFL against existing SOTA root cause localization methods, including ML/DL-based, and LLM-based approaches. As reported in Table~\ref{tab: method-comparison}, ThinkFL yields the highest MRR on all datasets except $\mathbf{B}$, where it is slightly outperformed by TraceRank. However, even in that case, ThinkFL remains highly competitive, with only a 9.63\% difference while achieving superior generalization across the remaining five datasets.

\begin{table}[htb]
	\setlength{\tabcolsep}{5pt}
	\centering
	\caption{Evaluation Results Compared with SOTA Methods}
	\label{tab: method-comparison}
	\begin{tabular}{c|c|cccccc}
		\toprule
		\multicolumn{2}{c|}{Method} & $\mathbf{A}$ & $\mathbf{B}$ & $\mathbf{\Gamma}$ & $\mathbf{\Delta}$ & $\mathbf{E}$ & $\mathbf{Z}$ \\
		\midrule
		\multirow{5}{*}{ML/DL-based} & CRISP & 8.27 & 20.13 & 18.13 & 17.34 & 31.08 & 17.14 \\
		~ & TraceConstract & 13.07 & 65.74 & \underline{58.55} & 2.48 & 33.77 & 8.15 \\
		~ & TraceRank & 6.26 & \textbf{76.76} & 34.41 & \underline{61.54} & \underline{35.79} & \underline{38.36} \\
		~ & MicroRank & 11.38 & 18.12 & 38.10 & 2.98 & 30.81 & 9.15 \\
		~ & RUN & 11.72 & 3.12 & 25.65 & 5.62 & 7.58 & 8.95 \\
		~ & MicroScope & 23.76 & 4.55 & 37.46 & 13.24 & 21.38 & 21.33 \\
		\midrule
		\multirow{2}{*}{LLM-based} & RCAgent & 17.59 & 20.20 & 23.95 & 14.64 & 12.65 & 16.35 \\
		~ & mABC & \underline{35.47} & 33.77 & 38.46 & 31.33 & 21.92 & 21.37 \\
		\midrule
		\multicolumn{2}{c|}{ThinkFL} & \textbf{54.44} & \underline{67.13} & \textbf{68.26} & \textbf{71.05} & \textbf{49.59} & \textbf{65.22} \\
		\bottomrule
	\end{tabular}
\end{table}

Compared to TraceRank, the strongest trace-based baseline, ThinkFL delivers an average improvement of 20.43\% in MRR, with particularly notable advantages on $\mathbf{A}$ (48.18 \%) and $\mathbf{\Gamma}$ (33.85 \%), demonstrating ThinkFL's ability to generalize across both noisy and structured traces. When compared to MicroScope, the best-performing metrics-based method, ThinkFL exhibits an even more dramatic improvement of 42.33\%, underscoring the limitations of relying solely on metrics in isolation. Likewise, ThinkFL exceeds the best-performing LLM-based method, mABC, by 32.23\%, showcasing the advantage of our design in integrating logs and metrics under a unified inference framework. It is also worth emphasizing that all LLM-based baselines are implemented on Qwen-2.5-Plus, a model with significantly more parameters than ThinkFL.

\subsection{Inference Time}

To further highlight the efficiency of ThinkFL, we conduct a comprehensive comparison of its inference time against SOTA LLMs. As presented in Table~\ref{tab: inference-time}, ThinkFL demonstrates remarkable speed, completing each failure localization task in an average of only 3.40 seconds across six representative datasets. This is approximately 673.11\% faster than the second-best model, Qwen-2.5-Plus, and outpaces heavier models such as Llama3.1-70B and DeepSeek-R1-Qwen by several orders of magnitude.

\begin{table}[htb]
	\setlength{\tabcolsep}{10pt}
	\centering
	\caption{Inference Speed Compared with SOTA Models (seconds/query)}
	\label{tab: inference-time}
	\begin{tabular}{c|cccccc}
		\toprule
		Model & $\mathbf{A}$ & $\mathbf{B}$ & $\mathbf{\Gamma}$ & $\mathbf{\Delta}$ & $\mathbf{E}$ & $\mathbf{Z}$ \\
		\midrule
		Claude-3.5-Sonnet & 90.38 & 51.30 & 89.45 & 44.00 & 29.83 & 29.40 \\
		Qwen-2.5-Plus & 30.87 & 23.98 & 26.05 & 19.55 & 27.57 & 29.85 \\
		Llama3.1-70B & 95.42 & 579.36 & 235.81 & 283.25 & 939.28 & 103.35 \\
		DeepSeek-R1-Qwen-32B & 194.81 & 154.14 & 149.31 & 143.47 & 234.43 & 333.42 \\
		Qwen-2.5-Max & 64.61 & 39.64 & 36.71 & 42.18 & 36.79 & 38.43 \\
		ThinkFL-3B & \textbf{3.51} & \textbf{3.30} & \textbf{3.30} & \textbf{3.54} & \textbf{3.38} & \textbf{3.39} \\
		\bottomrule
	\end{tabular}
\end{table}

This drastic reduction in inference time is not merely a numerical advantage—it translates directly into practical benefits in real-world system operations. In latency-sensitive environments where SREs must react swiftly to service disruptions, ThinkFL enables near-instantaneous root cause localization, thereby significantly shortening the mean time to recovery (MTTR). In contrast to models like Llama3.1-70B—which may require several minutes per failure case—ThinkFL reduces the latency to the order of seconds, making it far more practical for real-time troubleshooting scenarios.

Moreover, ThinkFL's speed advantage comes without sacrificing accuracy or robustness, proving that lightweight and targeted reasoning can outperform brute-force processing by large-scale models. Unlike general-purpose LLMs such as Llama3.1-70B, which suffer from massive parameter overhead and rely on overly generalized reasoning paths, ThinkFL’s domain-specialized design and efficient recursion-of-thought architecture enable it to operate effectively in low-latency regimes. This makes it particularly well-suited for production-scale environments where both time and resources are at a premium.

\subsection{Ablation Study}

To evaluate the effectiveness of our progressive multi-stage GRPO fine-tuning algorithm, we track the failure localization performance of ThinkFL after each stage on two representative backbone models: Llama3.2-3B and Qwen2.5-3B.

\begin{table}[htbp]
	\setlength{\tabcolsep}{4pt}
	\centering
	\caption{Ablation Study of Progressive GRPO (ThinkFL-Llama3.2-3B)}
	\label{tab: ablation-study-llama}
	\begin{tabular}{c|cccc|cccc|cccc}
		\toprule
		\multirow{2}{*}{Dataset} & \multicolumn{4}{c|}{Stage 1} & \multicolumn{4}{c|}{Stage 2} & \multicolumn{4}{c}{Stage 3} \\
		& R1 & R3 & R5 & MRR & R1 & R3 & R5 & MRR & R1 & R3 & R5 & MRR \\
		\midrule
		$\mathbf{A}$ 
		& 25.08 & 31.53 & 32.88 & 28.64
		& 22.37 & 50.85 & 54.24 & \underline{37.09}
		& 37.63 & 72.54 & 73.56 & \textbf{54.44} \\
		
		$\mathbf{B}$ 
		& 46.19 & 70.85 & 71.75 & 57.50
		& 69.51 & 70.85 & 73.09 & \underline{59.32}
		& 60.54 & 73.54 & 75.34 & \textbf{67.13} \\
		
		$\mathbf{\Gamma}$ 
		& 30.42 & 82.16 & 86.35 & 54.92
		& 61.08 & 63.83 & 64.67 &\underline{62.54}
		& 49.34 & 86.71 & 90.54 & \textbf{68.26} \\
		
		$\mathbf{\Delta}$ 
		& 46.19 & 76.64 & 81.86 & 62.01
		& 50.26 & 79.04 & 85.19 & \underline{64.24}
		& 60.90 & 80.60 & 83.73 & \textbf{71.05} \\
		
		$\mathbf{E}$ 
		& 32.16 & 35.46 & 35.68 & 33.86
		& 28.63 & 60.35 & 69.38 & \underline{45.34}
		& 33.04 & 69.38 & 76.43 & \textbf{51.48} \\
		
		$\mathbf{Z}$ 
		& 36.71 & 71.77 & 77.97 & 54.15
		& 40.89 & 74.43 & 79.37 & \underline{58.01}
		& 51.52 & 77.47 & 81.14 & \textbf{65.22} \\
		\bottomrule
	\end{tabular}
\end{table}

As shown in Table~\ref{tab: ablation-study-llama}, ThinkFL based on Llama3.2-3B already exhibits strong localization performance after Stage 1, with an average MRR of 48.51\% across six datasets. After Stage 2, the average MRR increases to 54.42\%, and further improves to 62.93\% after Stage 3, marking a relative improvement of 14.42\% from Stage 1 to Stage 3. Notably, Dataset $\mathbf{A}$ shows the most dramatic gain, with MRR rising from 28.64\% to 54.44\%—an increase of 25.80\%. These results suggest that GRPO effectively injects task-specific reasoning abilities in a staged manner.

To further verify the generalizability of GRPO, we apply the same fine-tuning stages to Qwen2.5-3B, and report the results in Table~\ref{tab: ablation-study-qwen}.

\begin{table}[htbp]
	\setlength{\tabcolsep}{4pt}
	\centering
	\caption{Ablation Study of Progressive GRPO (ThinkFL-Qwen2.5-3B)}
	\label{tab: ablation-study-qwen}
	\begin{tabular}{c|cccc|cccc|cccc}
		\toprule
		\multirow{2}{*}{Dataset} & \multicolumn{4}{c|}{Stage 1} & \multicolumn{4}{c|}{Stage 2} & \multicolumn{4}{c}{Stage 3} \\
		& R1 & R3 & R5 & MRR & R1 & R3 & R5 & MRR & R1 & R3 & R5 & MRR \\
		\midrule
		$\mathbf{A}$ 
		& 12.88 & 21.02 & 22.03 & 16.88
		& 16.61 & 38.64 & 40.68 & \underline{28.34}
		& 20.00 & 48.14 & 49.15 & \textbf{34.23} \\
		
		$\mathbf{B}$ 
		& 45.29 & 56.05 & 57.40 & 50.64
		& 37.67 & 63.68 & 69.06 & \underline{51.84}
		& 55.16 & 63.68 & 67.71 & \textbf{60.05} \\
		
		$\mathbf{\Gamma}$ 
		& 28.26 & 45.87 & 49.34 & 37.03
		& 27.19 & 59.04 & 64.55 & \underline{43.91}
		& 50.90 & 73.53 & 75.69 & \textbf{62.41} \\
		
		$\mathbf{\Delta}$ 
		& 38.06 & 55.16 & 61.63 & 47.59
		& 35.45 & 59.54 & 72.26 & \underline{50.85}
		& 55.16 & 67.88 & 74.97 & \textbf{63.04} \\
		
		$\mathbf{E}$ 
		& 15.64 & 36.78 & 40.75 & 26.23
		& 20.70 & 46.04 & 54.85 & \underline{35.68}
		& 26.65 & 49.56 & 54.85 & \textbf{39.04} \\
		
		$\mathbf{Z}$ 
		& 34.18 & 53.16 & 56.20 & 43.72
		& 30.63 & 59.62 & 67.09 & \underline{46.95}
		& 50.76 & 64.81 & 69.24 & \textbf{58.71} \\
		\bottomrule
	\end{tabular}
\end{table}

Here, ThinkFL-Qwen2.5-3B starts with an average MRR of 37.02\% in Stage 1, which rises to 42.93\% in Stage 2, and reaches 52.91\% in Stage 3—resulting in an overall 15.89\% relative improvement. The most impressive leap appears in Dataset $\mathbf{\Gamma}$, where MRR increases from 37.03\% to 62.41\%.

These consistent gains across both model architectures demonstrate that our GRPO fine-tuning strategy is both effective and generalizable. Each stage gradually introduces more complex reasoning skills—from basic grounding in Stage 1, to multi-hop reasoning in Stage 2, and finally to full recursive optimization in Stage 3—resulting in a model that not only localizes failures more accurately but does so in a more structured and explainable manner.

\subsection{Training Process Analysis}

Among the three stages in the progressive multi-stage GRPO fine-tuning algorithm, Stage 2 (Guided Exploration Augmentation) plays the most crucial role in enhancing the exploratory ability of lightweight LLMs. To better understand its internal dynamics, we conduct an in-depth analysis of key metrics throughout the training process.

\begin{figure}[htbp]
	\centering
	\subfigure[Reward]{
		\begin{minipage}{0.455\linewidth}
			\centering   
			\includegraphics[width=\textwidth]{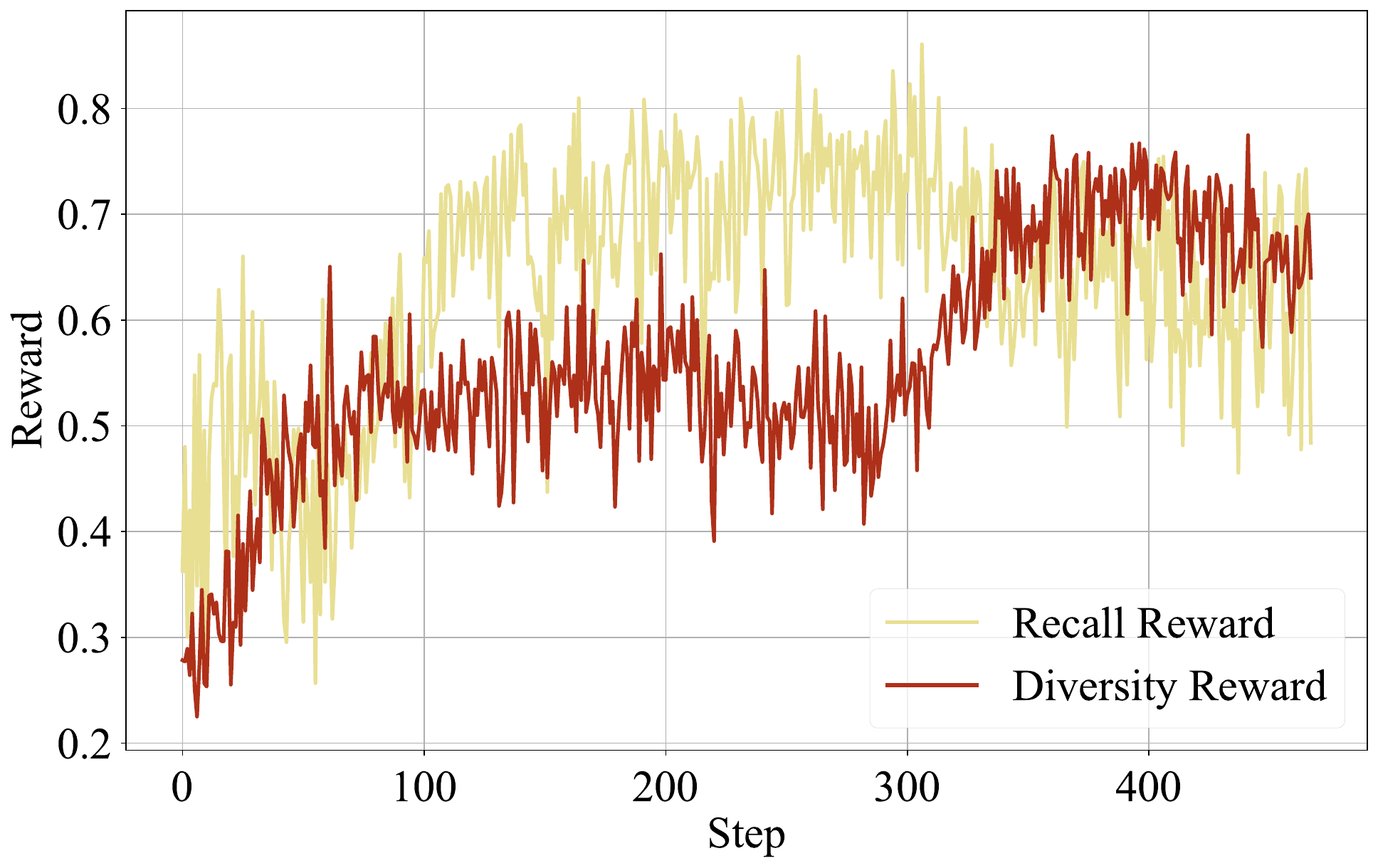}
			\label{fig:stage2-reward}
		\end{minipage}
	}
	\subfigure[Response Length]{
		\begin{minipage}{0.51\linewidth}
			\centering
			\includegraphics[width=\textwidth]{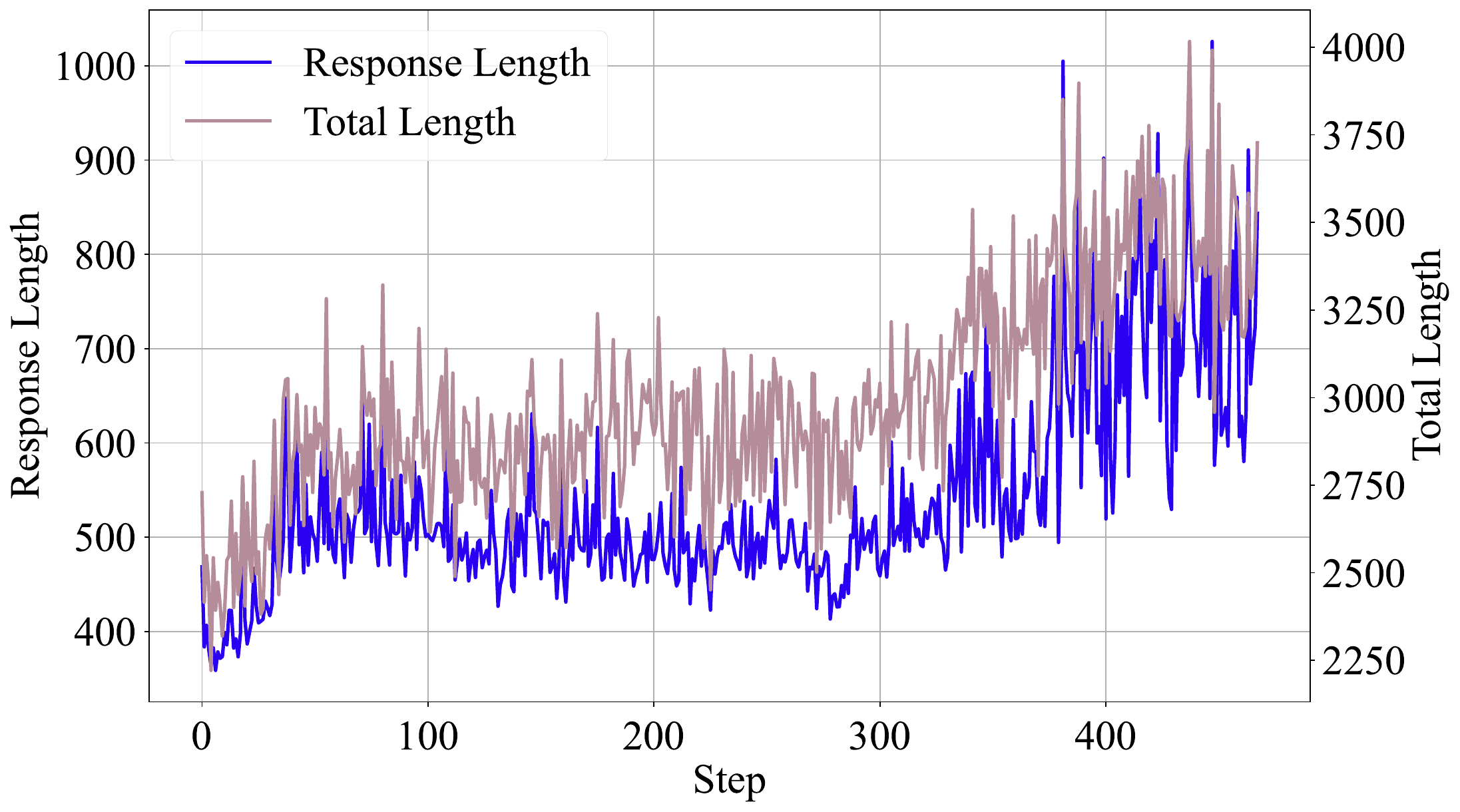}
			\label{fig:stage2-response-length}
		\end{minipage}
	}
	\caption{Training Effects on Stage2 (Guided Exploration Augmentation)}
	\label{fig:stage2-training}
\end{figure}

As shown in Figure~\ref{fig:stage2-training}, both recall reward and diversity reward exhibit distinct trends. In Figure~\ref{fig:stage2-reward}, the recall reward rapidly rises from 0.35 to 0.55 in the early steps, followed by a sharp drop back to the initial level. This suggests that the model initially tends to overfit conservative patterns. However, within the next 100 steps, the recall reward quickly rebounds and stabilizes at around 0.75, indicating that the model gradually learns to balance exploration and accurate identification.

In contrast, the diversity reward increases more steadily and with a slight delay. It climbs from 0.3 to approximately 0.5 in the first 300 steps, followed by a sudden surge to around 0.7 within the next 100 steps. This "slow-then-surge" pattern indicates that the model requires more time to effectively diversify its output, and that high diversity tends to emerge later in the training. Interestingly, this surge in diversity is not always beneficial—excessive diversity can dilute effective reasoning paths, which justifies the need for the final stage to constrain the output via ranking mechanisms.

We also track the length of the responses generated by ThinkFL at each step (response length), along with the combined length including tool call results (total length), as shown in Figure~\ref{fig:stage2-response-length}. Both metrics grow steadily in the first 300 steps—reaching approximately 500 and 3000 characters, respectively—before experiencing rapid increases. This trend coincides with the diversity spike, confirming that increased diversity correlates with longer, more exploratory outputs.

Overall, our analysis reveals that Stage 2 is essential for equipping the model with diverse exploratory abilities, while also highlighting the importance of proper constraint mechanisms in Stage 3 to harness and refine this exploration toward effective failure localization.

\subsection{LLM Backbone Comparison}

To further validate the generality and robustness of ThinkFL, we apply the same fine-tuning procedure across various LLM backbones, including Qwen2.5-0.5B, Qwen2.5-3B, Llama3-8B, and Llama3.2-3B. The final MRR results on six evaluation tasks are summarized in Table~\ref{tab: llm-backbone}.

\begin{table}[htbp]
	\setlength{\tabcolsep}{5pt}
	\centering
	\caption{Effectiveness of ThinkFL on Various LLMs}
	\label{tab: llm-backbone}
	\begin{tabular}{c|cccccc}
		\toprule
		Model & $\mathbf{A}$ & $\mathbf{B}$ & $\mathbf{\Gamma}$ & $\mathbf{\Delta}$ & $\mathbf{E}$ & $\mathbf{Z}$ \\
		\midrule
		ThinkFL-Qwen2.5-0.5B & 32.52 & 50.56 & 32.68 & 55.52 & 34.34 & 38.38 \\
		ThinkFL-Qwen2.5-3B & 34.23 & 60.05 & 62.41 & 63.04 & 39.04 & 58.71 \\
		ThinkFL-Llama3-8B & \underline{45.44} & \underline{60.14} & \underline{66.25} & \textbf{71.07} & \underline{50.15} & \underline{63.51} \\
		ThinkFL-Llama3.2-3B & \textbf{54.44} & \textbf{67.13} & \textbf{68.26} & \underline{71.05} & \textbf{51.48} & \textbf{65.22} \\
		\bottomrule
	\end{tabular}
\end{table}

From the results, we observe that while scaling up model size often brings performance gains, model size alone is not the sole determinant of effectiveness. For instance, despite being smaller, ThinkFL-Llama3.2-3B surpasses the larger Llama3-8B, highlighting that architectural refinements and instruction-tuning quality can outweigh sheer parameter count.

Nevertheless, a general scaling trend is evident when comparing Qwen2.5-0.5B to Qwen2.5-3B. The average MRR increases by 12.24\% (from approximately 40.67\% to 52.91\%), with the most noticeable improvements in tasks $\Gamma$ and $\Delta$, where MRR rises from 32.68\% to 62.41\% and from 55.52\% to 63.04\%, respectively. These results indicate that models around the 3B scale strike a strong balance between accuracy and parameter efficiency for failure localization.

Models from the Llama family further demonstrate robust performance. ThinkFL-Llama3-8B achieves an average MRR of approximately 59.43\% across all tasks and stands out in tasks $\Gamma$ and $\mathbf{Z}$, with scores of 66.25\% and 63.51\%—outperforming Qwen2.5-3B by 3.84\% and 4.8\%, respectively. This suggests that Llama-based models possess stronger reasoning and multi-step planning capabilities.

Finally, ThinkFL-Llama3.2-3B achieves the highest overall performance with an average MRR of approximately 62.93\%, outperforming its larger counterpart Llama3-8B by 3.50\%. Notably, it delivers significant gains on task $\mathbf{A}$ (54.44\% vs. 45.44\%) and task $\mathbf{E}$ (51.48\% vs. 50.15\%). This exception underscores that model architecture refinements and instruction-tuning quality, as exemplified by Llama3.2, can play a decisive role. Such findings make Llama3.2-3B particularly attractive for real-world deployments, where computational efficiency is as critical as accuracy.

\subsection{Case Study}

To better understand why ThinkFL achieves improved failure localization performance after self-refinement, we present a case study comparing its reasoning paths before and after the full progressive training pipeline.

\begin{figure}[htbp] 
	\centering 
	\includegraphics[width=1\linewidth]{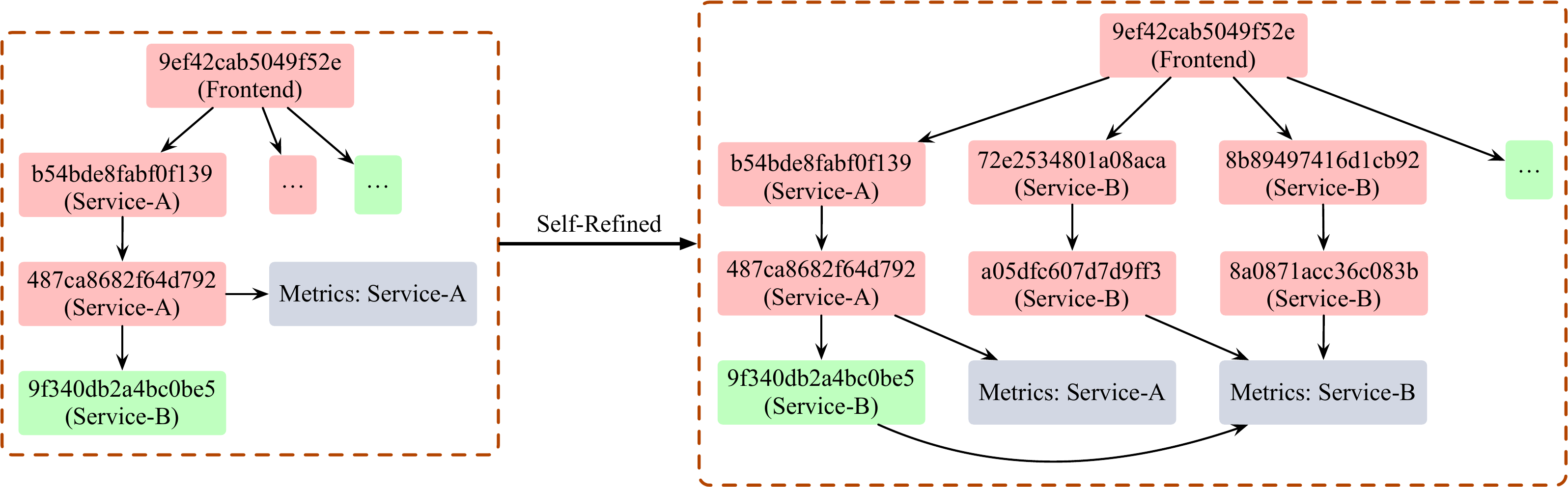} 
	\caption{Case Study: Reasoning Path Comparison before and after Self-Refinement} 
	\label{fig: case-study} 
\end{figure}

As illustrated in Figure~\ref{fig: case-study}, we examine two reasoning trajectories generated by ThinkFL for the same failure event with the root trace ID \texttt{9ef42cab5049f52e}. The left side depicts the output of ThinkFL trained only with Stage 1 (format-aware policy priming), while the right shows the behavior after completing the full self-refinement process (Stage 2 and Stage 3).

In the Stage 1 setting, ThinkFL detects an initial abnormal sub-call \texttt{b54bde8fabf0f139} and proceeds to follow its downstream chain. It then identifies another abnormal call \texttt{487ca8682f64d792}, but eventually reaches a normal sub-call \texttt{9f340db2a4bc0be5}. Based on this limited exploration, it concludes that the root cause likely lies in \texttt{Service-A}, to which \texttt{487ca8682f64d792} belongs.

In contrast, the self-refined ThinkFL exhibits significantly broader exploration behavior. Instead of strictly following a single path, it also investigates two sibling sub-calls: \texttt{72e2534801a08aca} and \texttt{8b89497416d1cb92}, both of which belong to \texttt{Service-B}. Since these traces form abnormal chains within the same service, ThinkFL aggregates information from both \texttt{Service-A} and \texttt{Service-B}, including relevant metrics, to form a more holistic judgment. It ultimately attributes the failure to \texttt{Service-B} — the ground-truth root cause.

This comparison clearly demonstrates ThinkFL’s enhanced capability to localize failures more accurately after undergoing self-refinement. The model’s ability to escape early-stage greedy search and explore alternative hypotheses not only improves the reasoning depth but also leads to better coverage of plausible failure paths. It is worth noting that this ability to perform diversified and metric-grounded exploration is also one of the key reasons why ThinkFL outperforms even significantly larger models such as Claude-3.5-Sonnet in our evaluations. 

\begin{figure}[htbp]
	\centering
	\includegraphics[width=1\linewidth]{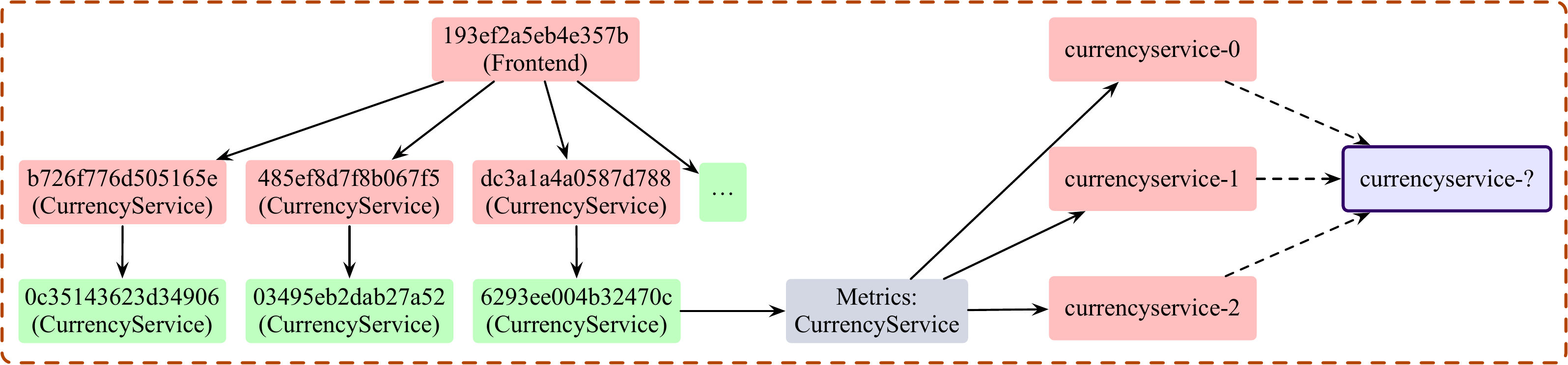}
	\caption{Case Study of a Failed Failure Localization}
	\label{fig:case-study-fail}
\end{figure}

Although ThinkFL has made significant progress after Progressive Multi-Stage GRPO Fine-Tuning, there still exist cases where it fails to accurately localize failures. As shown in Figure~\ref{fig:case-study-fail}, this example involves a relatively simple trace, where all anomalies can be traced back to the \texttt{CurrencyService}. Ideally, the model only needs to query the metrics of \texttt{CurrencyService} to confirm the root cause. However, upon inspection, all three related pods—\texttt{currencyservice-0}, \texttt{currencyservice-1}, and \texttt{currencyservice-2}—exhibited abnormal fluctuations. In this situation, ThinkFL found it difficult to interpret the metrics and distinguish which pod was the true source of the failure. Across multiple reasoning attempts, the model produced different predictions, with most of them being incorrect.

This example indicates that ThinkFL’s current limitation lies not in its ability to follow structured reasoning, but in its insufficient understanding of performance metrics when multiple components exhibit correlated anomalies. Enhancing the model’s capability to interpret metric and differentiate between correlated and causal anomalies will be an important avenue for future improvement.

\subsection{Hyperparameter Analysis}

Among all hyperparameters, the most critical one is the maximum dialogue depth $D_{\text{max}}$, which controls the depth of the model's exploration process. To understand its impact, we conduct experiments on the $\mathbf{\Delta}$ dataset — a medium-difficulty benchmark — by varying $D_{\text{max}}$ and evaluating its effect on failure localization performance.

\begin{figure}[htbp]
	\centering
	\subfigure[ThinkFL-Llama3.2-3B]{
		\begin{minipage}{0.48\linewidth}
			\centering   
			\includegraphics[width=\textwidth]{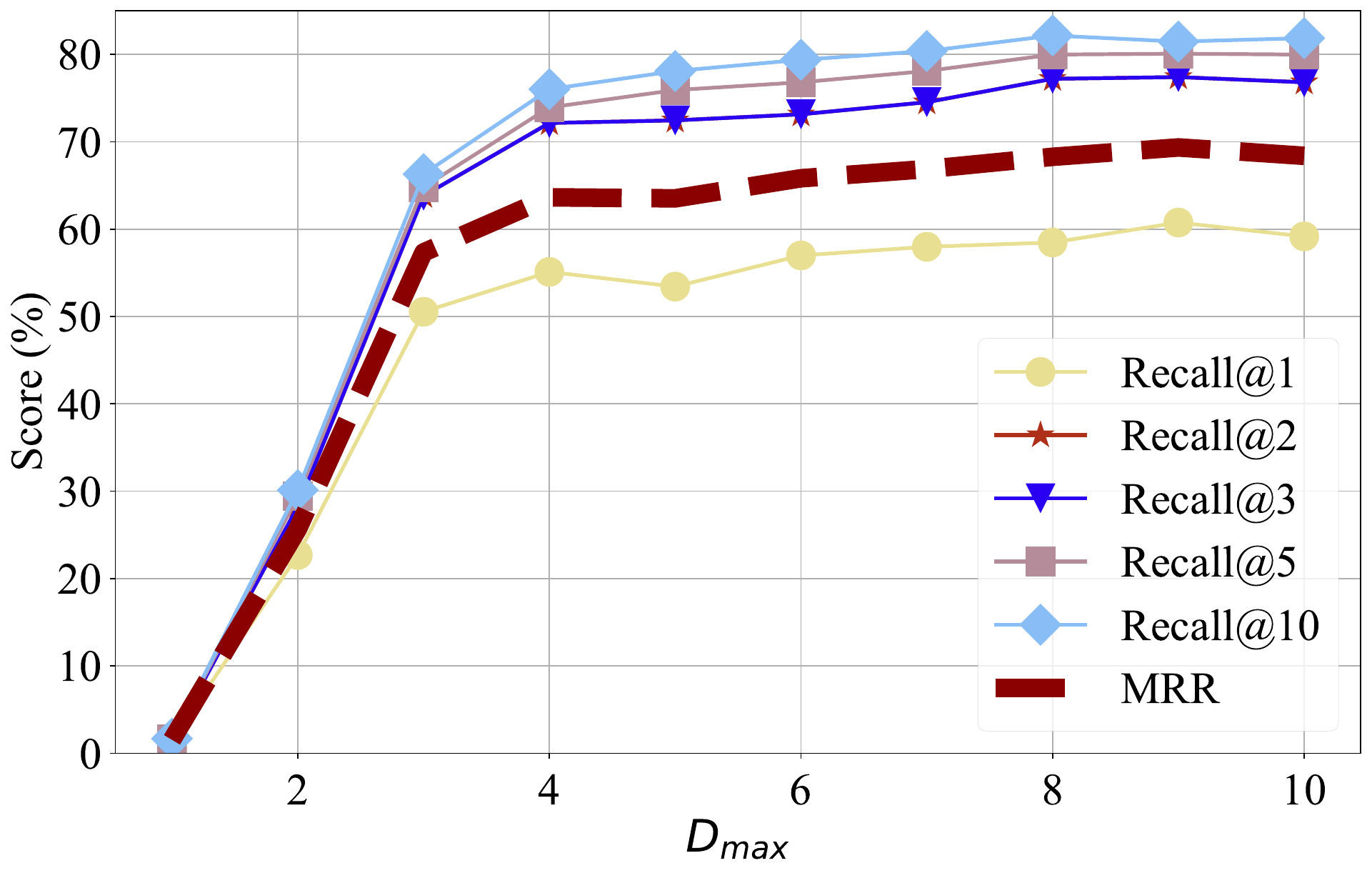}
			\label{fig:params-loop-time-llama3.2-3b}
		\end{minipage}
	}
	\subfigure[ThinkFL-Qwen2.5-3B]{
		\begin{minipage}{0.48\linewidth}
			\centering
			\includegraphics[width=\textwidth]{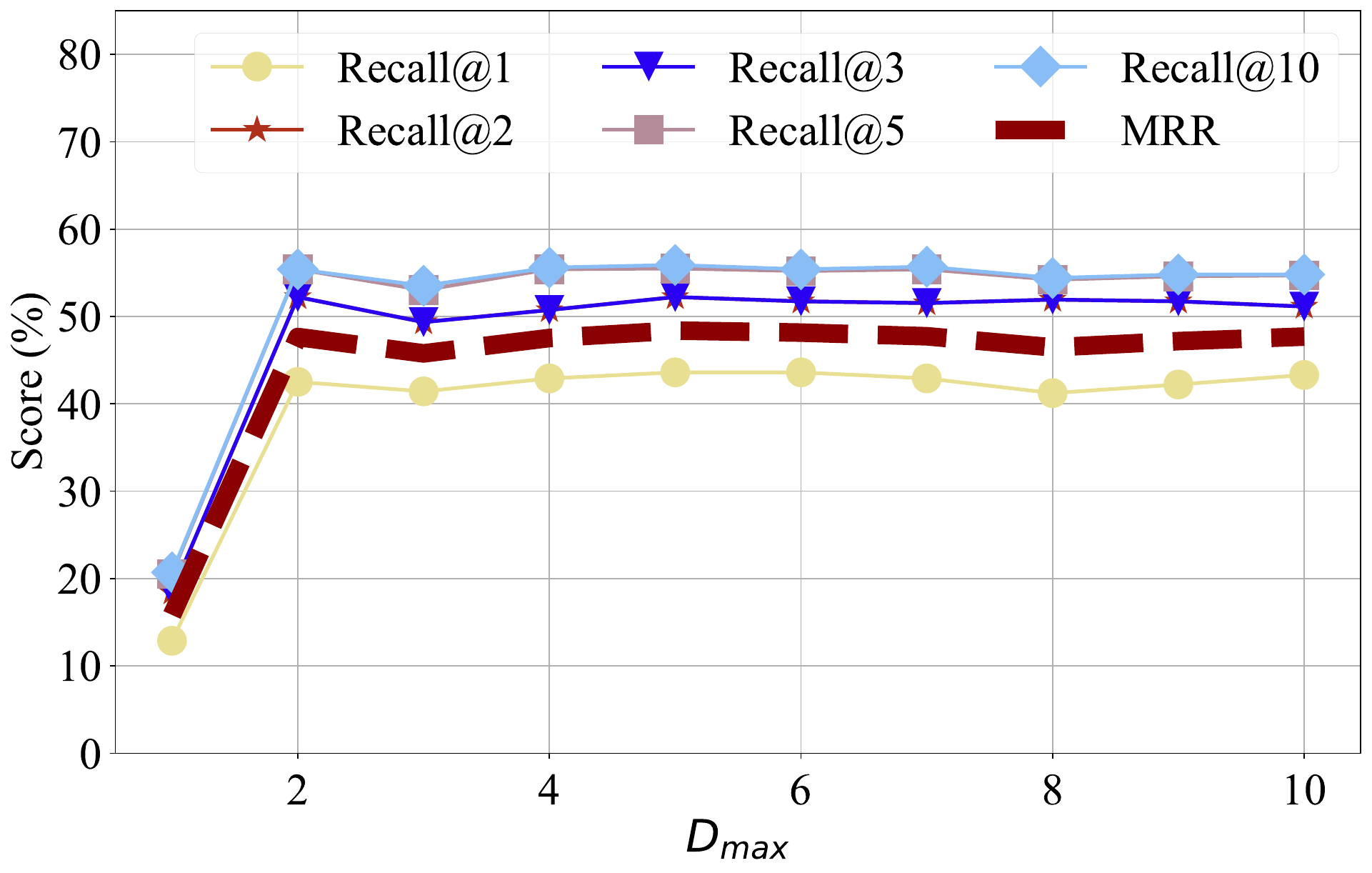}
			\label{fig:params-loop-time-qwen2.5-3b}
		\end{minipage}
	}
	\caption{Hyperparameter Experiment of $D_{max}$}
	\label{fig:params-loop-time}
\end{figure}

As shown in Figure~\ref{fig:params-loop-time}, increasing $D_{\text{max}}$ generally leads to improved failure localization performance, but the improvement eventually plateaus. 

For ThinkFL-Llama3.2-3B (Figure~\ref{fig:params-loop-time-llama3.2-3b}), we observe a substantial performance gain when increasing $D_{\text{max}}$ from 1 to 4, achieving a 62.09\% improvement in MRR. Beyond $D_{\text{max}}=4$, the gain continues but at a slower rate, with an additional 5.71\% improvement from $D_{\text{max}}=4$ to $10$. This suggests that deeper dialogue depths can still contribute marginal improvements, especially for more capable backbones like Llama3.2-3B, which benefit from extended reasoning chains.

In contrast, ThinkFL-Qwen2.5-3B (Figure~\ref{fig:params-loop-time-qwen2.5-3b}) reaches its peak performance much earlier. The MRR increases by 31.81\% from $D_{\text{max}}=1$ to $2$, but further increases in dialogue depth offer little to no improvement. This indicates that Qwen2.5-3B may struggle to fully utilize deeper multi-hop reasoning, possibly due to its architectural or training limitations.

These results suggest that while increasing $D_{\text{max}}$ is generally beneficial, the optimal value depends heavily on the backbone's reasoning capacity. For lighter or less structured models, shallower dialogues may suffice, whereas more powerful models can leverage longer reasoning paths to refine their localization hypotheses more effectively.

\subsection{Generalizability Evaluation}

To evaluate ThinkFL's generalizability to other microservice systems, we applied it to the Train-Ticket system\footnote{https://github.com/FudanSELab/train-ticket}, which we manually set up with 41 microservices and injected 10 types of faults to simulate failures. Notably, ThinkFL was used directly without any retraining on this system.

\begin{table}[htbp]
	\setlength{\tabcolsep}{5pt}
	\centering
	\caption{Failure Localization Results in Train-Ticket}
	\label{tab: generalizability-evaluation}
	\begin{tabular}{c|cccccc}
		\toprule
		Method & $Recall@1$ & $Recall@2$ & $Recall@3$ & $Recall@5$ & $Recall@10$ & $MRR$ \\
		\midrule
		RCAgent & 41.33 & 51.75 & 58.13 & 64.97 & 73.17 & 59.18 \\
		mABC & 44.89 & 54.97 & 59.88 & 68.12 & 75.32 & 61.33 \\
		\textbf{ThinkFL} (\textit{ours}) & 61.23 & 71.30 & 73.26 & 81.25 & 87.67 & 76.32 \\
		\bottomrule
	\end{tabular}
\end{table}

As shown in Table~\ref{tab: generalizability-evaluation}, ThinkFL substantially outperforms the baselines. For instance, its $Recall@1$ of 61.23\% is 16.34 points higher than mABC (44.89\%) and 19.90 points higher than RCAgent (41.33\%). Similarly, the mean reciprocal rank (MRR) of ThinkFL reaches 76.32, which represents an improvement of 15.0–17.1 points over the baselines. Across other metrics ($Recall@2$, $Recall@3$, $Recall@5$, $Recall@10$), ThinkFL consistently shows gains of 10–15 percentage points, highlighting its ability to accurately identify root causes in an unseen microservice system.

These results indicate that the model trained via Progressive Multi-Stage GRPO Fine-Tuning generalizes effectively to an environment with different services, architectures, and fault patterns. Importantly, ThinkFL achieves this performance without any system-specific retraining, demonstrating its robustness and practical applicability in diverse operational settings.

\section{Discussion}

\subsection{Distinction Between Research Evaluation and Practical Usage}

Our evaluation isolates the technical capabilities of ThinkFL, whereas real-world failure diagnosis is fundamentally a human-in-the-loop process. In practice, SREs interpret model outputs, validate evidence, and decide on remediation actions. While our experiments do not measure these human interactions, they assess the core technical abilities required for practical deployment: correct tool invocation, structured reasoning generation, and accurate root-cause identification at multiple granularities.

The goal of this work is the research contribution—advancing the reasoning ability of lightweight LLMs through Progressive Multi-Stage GRPO Fine-Tuning, including the Recursion-of-Thought framework and format-aware policy priming. Accordingly, our evaluation focuses on controlled technical metrics rather than end-to-end incident-management workflows.

In operational use, ThinkFL is intended as an assistive system: it produces interpretable reasoning traces and ranked candidate root causes that SREs can verify and act upon. The structured outputs provide transparent evidence that supports faster confirmation and remediation. To partially verify this practical utility, we conducted a small-scale, informal study by presenting selected interpretable reasoning paths and predictions to professional developers and SREs. They reported that these interpretable traces improved their understanding of the diagnostic process and helped them reach resolutions more quickly. This preliminary check suggests practical promise, although a systematic evaluation with a larger SRE cohort and real incident-management scenarios remains future work due to resource constraints.

\subsection{Dependence on Microservice Characteristics and Potential Extension}

ThinkFL is designed around microservice observability, where spans, traces, and metrics provide rich structured signals for reasoning. These signals enable ThinkFL to trace request propagation, correlate anomalies with service- or pod-level behavior, and generate interpretable Recursion-of-Thought (RoT) reasoning chains. Thus, its current performance naturally benefits from characteristics inherent to microservice architectures.

Our evaluation on the Train-Ticket benchmark (41 services, 10 injected failures) further reflects this dependency. ThinkFL performs well on this dataset partly because its failures are simpler and more localized than those in AIOPS 2022. This demonstrates both its generalizability across microservice systems and its sensitivity to system scale and fault complexity.

Nevertheless, ThinkFL is not fundamentally limited to microservices. Its design consists of three modular layers—(i) a trace abstraction layer, (ii) a metrics integration layer, and (iii) a reasoning/localization layer—only the first two of which are domain-specific. By redefining the trace abstraction and mapping metrics to domain-specific signals, ThinkFL’s reasoning layer can operate unchanged.

In summary, ThinkFL’s success stems from microservice-aligned observability, but its modular structure enables extension to other distributed systems as long as comparable structured traces and metrics can be provided.

\subsection{Handling Context Window Limitations in Recursive Analysis}

A natural concern regarding ThinkFL lies in whether recursive tool invocations and long traces might cause the accumulated context to exceed the LLM’s window size. We acknowledge that such situations could theoretically occur; however, our framework was explicitly designed with this constraint in mind. Instead of passing the entire raw trace into the model at once, ThinkFL performs iterative querying: at each step, only the necessary fragment of the trace is retrieved and fed into the model. This design keeps the effective context size small and manageable.

In practice, the traces in the AIOPS 2022 dataset contain at most five levels of depth. Consequently, during our experiments, we did not encounter cases where the accumulated trace length exceeded the model’s context window. Nevertheless, to further safeguard against excessive recursive queries, we impose a maximum reasoning depth of 20 tool-invocation rounds. If this limit is reached, ThinkFL terminates the exploration process and produces a final root cause judgment based on the available information. This mechanism ensures that ThinkFL remains both context-efficient and robust, even under scenarios with potentially large or repetitive traces.

\subsection{Practical Implications for SREs}

We acknowledge the concern regarding the effort and resources required to deploy ThinkFL. In practice, training and inference only require lightweight resources—e.g., several consumer-grade GPUs such as NVIDIA RTX 4090 are sufficient for the 3B-parameter backbone—making the computational cost manageable compared to large-scale LLMs.

Before training ThinkFL, we manually verified that advanced models (e.g., Claude-3.5) can generate correct tool-invocation sequences and reasoning chains for some failure localization cases. While these models occasionally produce incomplete reasoning or incorrect conclusions, we selected only correctly solved cases during the supervised fine-tuning (SFT) phase. As a result, ThinkFL can learn accurate reasoning patterns from advanced models, reducing the need for extensive manual annotations by SREs.

From the SRE perspective, ThinkFL does not introduce heavy additional workload. Recording new cases for incremental improvement requires only lightweight annotations (e.g., timestamp + root cause), which aligns with routine incident reporting. Moreover, since ThinkFL produces interpretable reasoning paths, its outputs remain useful even when imperfect, helping to reduce investigation overhead and accelerate failure triage. Overall, the combination of manageable GPU cost, minimal annotation requirements, and training data largely sourced from verified advanced model outputs makes the solution practically satisfactory for operators.

\subsection{Limitations and Threats to Validity}

Despite the promising results demonstrated by ThinkFL, several limitations and potential threats to validity should be acknowledged:

\textbf{Implementation of Baselines.} While we directly used publicly available source code for CRISP, MicroRank, and mABC, other baseline methods were re-implemented based on the descriptions provided in their respective papers. To ensure faithful reproduction, we carefully followed the algorithmic details, hyperparameter settings, and experimental protocols whenever available. For methods lacking released code or complete configuration details, we acknowledge that our implementations may deviate slightly from the originals. To mitigate this risk, multiple members of our research team independently reviewed and cross-checked the re-implemented code to ensure correctness and consistency.

\textbf{Reliance on OpenRLHF.} Our method is implemented based on the OpenRLHF framework, which provides a general foundation for reinforcement learning with LLMs. While we made necessary adaptations for our setting, there is a possibility that some default configurations in the framework—if not fully aligned with our objectives—could affect the final training performance. To mitigate this, we invited multiple domain experts to review our implementation in detail and ensure correctness and consistency.

\textbf{Simplified Failure Definition.} ThinkFL focuses exclusively on failure localization and does not include an anomaly detection component. Therefore, our definition of a failure is simplified: any request whose latency exceeds 100 times the normal threshold is considered anomalous. While this heuristic helps reduce false positives and ensures consistency in our evaluation, it may miss subtler forms of failure or introduce labeling noise.

\section{Related Work}

\subsection{Root Cause Localization}

Root cause localization is a fundamental component of failure management, aiming to identify the services, components, or operations responsible for system anomalies. Existing approaches can be broadly categorized into two types: ML/DL-based methods and LLM-based methods.

\vspace{0.5em} \noindent\textbf{ML/DL-based Methods.}
Traditional machine learning and deep learning approaches form the foundation of prior research on failure localization. These methods typically learn correlations, dependencies, or causal relationships from system telemetry—such as KPIs, logs, and traces—to identify abnormal components.

A large body of work focuses on analyzing system-level KPIs. Approaches such as LOUD~\cite{mariani2018localizing}, AID~\cite{yang2021aid}, and CloudScout~\cite{yin2016cloudscout} construct correlation graphs or compute similarity measures over KPIs (e.g., CPU, memory, disk I/O) to infer faulty services. More advanced techniques integrate causality analysis or forecasting, including CIRCA~\cite{li2022causal}, RUN~\cite{lin2024root}, and KPIRoot~\cite{gu2024kpiroot}, which leverage Bayesian networks, neural Granger causality, or symbolic KPI encoding for improved root cause ranking. Other methods target high-dimensional KPI attributes, such as CMMD~\cite{yan2022cmmd}, HALO~\cite{zhang2021halo}, iDice~\cite{lin2016idice}, and MID~\cite{gu2020efficient}, which use graph attention, conditional entropy, statistical scoring, or meta-heuristic search to identify fault-indicating metric combinations.

In parallel, another line of work analyzes distributed traces to capture end-to-end service interactions. Early efforts~\cite{li2022enjoy, luo2021characterizing} demonstrated the diagnostic value of trace structures. Subsequent learning-based techniques—such as MEPFL~\cite{zhou2019latent}, Seer~\cite{gan2019seer}, TraceAnomaly~\cite{liu2020unsupervised}, and Sage~\cite{gan2021sage}—apply supervised or unsupervised models, including CNNs, LSTMs, and GNNs, to detect anomalies and infer faulty components from trace patterns.

To improve interpretability and reduce reliance on large training datasets, spectrum-based methods have also been widely explored. MicroRank~\cite{yu2021microrank}, TraceRank~\cite{yu2023tracerank}, and the method of Li et al.~\cite{li2021practical} compute suspiciousness scores based on trace evidence, while TraceConstract~\cite{zhang2024trace} integrates sequence modeling with contrast pattern mining to localize complex multi-dimensional root causes.

Several human-in-the-loop approaches further incorporate SRE feedback to guide or refine the localization process. HRLHF~\cite{wang2023root}, HiLogX~\cite{jia2024hilogx}, and HiLog~\cite{jia2022augmenting} leverage expert annotations or interaction signals to improve dependency inference or log-based anomaly detection.

Despite their diversity, these ML/DL-based techniques share common limitations: they rely on predefined causal graphs, fixed statistical assumptions, or models trained on historical data. As a result, they often require substantial feature engineering, retraining, or manual updates when system behaviors evolve, and may struggle to generalize to previously unseen failure patterns.

\vspace{0.5em} \noindent\textbf{LLM-based Methods.}
Recent work applies large language models to failure localization, leveraging their ability to conduct flexible and interpretable reasoning over heterogeneous operational data. Existing approaches can be broadly grouped into two categories: tool-augmented single-agent systems and LLM-based multi-agent frameworks.

The first category consists of tool-augmented single-agent systems, where an LLM is equipped with domain-specific tools that preprocess, filter, or reconstruct diagnostic context before reasoning. Representative examples include RCAgent~\cite{wang2024rcagent}, which integrates log and code analysis tools to support tasks such as predicting root causes and gathering evidence; COCA~\cite{li2025coca}, which enhances issue-report analysis by extracting relevant code snippets and reconstructing execution paths; and TAMO~\cite{wang2025tamo}, which unifies multi-modal operational signals into time-aligned representations and incorporates specialized tools for component-level localization.

The second category comprises LLM-based multi-agent frameworks, where multiple agents collaborate through structured workflows. Flow-of-Action~\cite{pei2025flow} introduces an SOP-enhanced architecture in which procedural knowledge constrains LLM decisions at key junctures, ensuring a more reliable diagnostic trajectory. mABC~\cite{zhang2024mabc} coordinates agents via workflow rules and voting-based consensus, while KnowledgeMind~\cite{ren2025multi} employs Monte Carlo Tree Search and a knowledge-base reward mechanism to support standardized, service-by-service reasoning.

Although these approaches improve interpretability and incorporate valuable domain knowledge, they lack the self-iterative reasoning capability of ThinkFL and often struggle to precisely localize the true root-cause components when diagnostic context must be dynamically constructed rather than statically preprocessed.

\subsection{LLM-based Failure Management}

Large language models, with their strong semantic comprehension and reasoning capabilities, have shown great potential in advancing various aspects of failure management~\cite{zhang2025survey}. Recent studies have leveraged LLMs to improve anomaly detection, failure diagnosis, and automated remediation. These efforts can be broadly classified into three categories: foundation models, fine-tuning-based approaches, and prompt-based methods.

\vspace{0.5em}
\noindent\textbf{Foundation Models.}
Several studies aim to build foundation models specifically designed for time-series or log data. Lag-Llama~\cite{rasul2023lag} pretrains a decoder-only transformer for univariate forecasting by modeling lag-based covariates. Timer~\cite{liu2024timer} develops a GPT-style model that unifies forecasting, imputation, and anomaly detection for long-term sequences. TimesFM~\cite{das2024decoder} proposes a patched-decoder attention architecture to handle diverse forecasting lengths and granularities. For command-line automation, ShellGPT~\cite{shi2023shellgpt} incorporates shell scripting semantics into the GPT framework. TimeGPT~\cite{liao2024timegpt} and SimMTM~\cite{dong2024simmtm} adopt encoder-decoder structures for reconstructing time-series patterns from massive datasets. In the log domain, PreLog~\cite{le2024prelog} uses hierarchical objectives to enable pretraining for both log parsing and anomaly detection. KAD-Disformer~\cite{yu2024pre} further disentangles temporal and spatial dynamics to enhance detection performance in multivariate metrics data.

\vspace{0.5em}
\noindent\textbf{Fine-Tuning-Based Approaches.}
Fine-tuning general-purpose LLMs on domain-specific tasks is another common strategy. UniTime~\cite{liu2024unitime} fine-tunes GPT-2 for unified time-series forecasting across domains. AnomalyLLM~\cite{liu2024anomalyllm} applies contrastive fine-tuning on LLaMA2-7B for graph-based anomaly detection. PromptCast~\cite{xue2023promptcast} adapts T5 and BART via task-oriented fine-tuning to support text-visible time-series forecasting. RAG4ITOps~\cite{zhang2024rag4itops} applies retrieval-augmented fine-tuning on Qwen-14B for interactive failure diagnosis. OWL~\cite{guo2023owl} fine-tunes LLaMA2-13B using adapter modules for assisted troubleshooting and log anomaly detection. Similarly, LLM4TS~\cite{chang2023llm4ts} freezes most GPT-2 parameters while tuning a small subset for forecasting tasks. LogLM~\cite{liu2024loglm} introduces instruction-based fine-tuning on LLaMA2-7B to support multi-task log analytics, including parsing and detection.

\vspace{0.5em}
\noindent\textbf{Prompt-Based Methods.}
Prompt-based methods avoid the cost of full fine-tuning by leveraging LLMs through instruction design. RCACopilot~\cite{chen2024automatic} and Xpert~\cite{jiang2024xpert} use in-context learning (ICL) to structure multi-step diagnostic reasoning. LasRCA~\cite{han2024potential} employs one-shot prompting with auxiliary classifiers to improve anomaly detection on metrics. LSTPrompt~\cite{liu2024lstprompt} decomposes prediction into short- and long-term subtasks using chain-of-thought (CoT) prompting. LogGPT~\cite{liu2024interpretable} enhances log anomaly detection via step-wise CoT generation. LM-PACE~\cite{zhang2024lm} further strengthens diagnostic interpretability by prompting GPT-4 to analyze incident reports. Retrieval-augmented methods such as RAGLog~\cite{pan2023raglog}, LogRAG~\cite{zhang2024lograg} enhance log-based anomaly detection by retrieving and incorporating historical case logs into the prompt context. ThinkRemed~\cite{zhang2025microremed} employs a multi-agent framework to address microservice remediation.

\section{Conclusion}

In this paper, we address the limitations of existing LLM-based failure localization methods, such as rigid invocation workflows and resource-intensive inference. To overcome these challenges, we explore the use of reinforcement fine-tuning to endow lightweight LLMs with reasoning and self-refinement capabilities, thereby enhancing the cost-effectiveness and adaptability of LLM-based failure localization in real-world applications.

To this end, we first conduct an empirical study and identify three core capabilities essential for effective failure localization. Based on these insights, we propose a progressive multi-stage GRPO fine-tuning framework, which incorporates a multi-factor failure localization grader and a resursion-of-thought actor. The resulting model, ThinkFL, not only surpasses all existing state-of-the-art LLMs and methods in localization accuracy, but also significantly reduces localization latency from minutes to seconds.

In future work, we plan to investigate whether such reasoning-capable models can be extended beyond accurate failure localization to achieve automated recovery in runtime microservice systems.

\section{Data Availability}

The training code, which is based on OpenRLHF~\cite{hu2024openrlhf}, is publicly available at \url{https://github.com/LLM4AIOps/OpenRLHF-ThinkFL}. In addition, we implemented an alternative training version based on AgentEvolver~\cite{zhai2025agentevolver}, which is available at \url{https://github.com/modelscope/AgentEvolver}. The fine-tuned model weights are publicly released on ModelScope\footnote{ModelScope links: \url{https://modelscope.cn/models/ZhangLingzhe/ThinkFL-Qwen2.5-0.5B}, \url{https://modelscope.cn/models/ZhangLingzhe/ThinkFL-Qwen2.5-3B}, \url{https://modelscope.cn/models/ZhangLingzhe/ThinkFL-Llama3.2-3B}, \url{https://modelscope.cn/models/ZhangLingzhe/ThinkFL-Llama3-8B}.} and Hugging Face\footnote{ Hugging Face links: \url{https://huggingface.co/Zhang-Lingzhe/ThinkFL-Qwen2.5-0.5B}, \url{https://huggingface.co/Zhang-Lingzhe/ThinkFL-Qwen2.5-3B}, \url{https://huggingface.co/Zhang-Lingzhe/ThinkFL-Llama3.2-3B}, \url{https://huggingface.co/Zhang-Lingzhe/ThinkFL-Llama3-8B}.}.

\begin{acks}
 This work is supported by Key-Area Research and Development Program of Guangdong Province, China (NO.2020B010164003).
\end{acks}

\bibliographystyle{ACM-Reference-Format}
\bibliography{sample-base}

\clearpage
\appendix

\begin{figure}[h]
	\centering
	\begin{tcolorbox}[colback=gray!10,
		colframe=black,
		width=\linewidth,
		arc=1mm, auto outer arc,
		boxrule=0.5pt,
		top=5pt, bottom=5pt, left=5pt, right=5pt
		]
		
		\footnotesize
		
		\textbf{[System]} You are a software operations engineer. Your task is to systematically diagnose and identify the root cause of software failures. You have the following tools: search\_traces, search\_fluctuating\_metrics, print\_results.
		
		\vspace{0.2cm}
		\hrule
		\vspace{0.2cm}
		
		\noindent\textbf{[User]} Please read the following root trace and identify corresponding root cause service. \\
		\texttt{timestamp: 1647753157852, cmdb\_id: frontend2-0, span\_id: 0a81f08fc9b7dc5d, duration: 29982953 \textmu\textrm{s}}
		
		\vspace{0.2cm}
		\hrule
		\vspace{0.2cm}
		
		\noindent\textbf{[Tool Call]} \texttt{search\_traces(parent\_span\_id="0a81f08fc9b7dc5d")}
		
		\noindent\textbf{[Tool Response]}
		
		\noindent
		\begin{tabular}{llrr}
			cmdb\_id & opeartion & duration & span\_id \\
			frontend2-0 & RecommendationService/ListRecommendations & 29,892,726 \textmu\textrm{s} & 9063994c3450e63a\\
			frontend2-0 & CurrencyService/GetSupportedCurrencies & 1,836 \textmu\textrm{s} & 1c7b764e0a11a132\\
			frontend2-0 & CartService/GetCart & 1,995 \textmu\textrm{s} & 0dc30c7af1adbe40\\
			frontend2-0 &AdService/GetAds & 4,516 \textmu\textrm{s} & a383db1c6acc90a5\\
			frontend2-0 &ProductCatalogService/GetProduct & 41813 \textmu\textrm{s} & ceed3378cb0c32f9\\
			... & ... & ... & ...
		\end{tabular}
		
		\hrule
		\vspace{0.2cm}
		
		\noindent\textbf{[Tool Call]} \texttt{search\_traces(parent\_span\_id="9063994c3450e63a")}
		
		\noindent\textbf{[Tool Response]}
		
		\noindent
		\begin{tabular}{llrr}
			cmdb\_id & opeartion & duration & span\_id \\
			recommendationservice2-0 & RecommendationService/List & 29889199 \textmu\textrm{s} & eedd72a7aaa04418\\
		\end{tabular}
		
		\hrule
		\vspace{0.2cm}
		
		\noindent\textbf{[Tool Call]} \texttt{search\_traces(parent\_span\_id="eedd72a7aaa04418")}
		
		\noindent\textbf{[Tool Response]}
		
		\noindent
		\begin{tabular}{llrr}
			cmdb\_id & opeartion & duration & span\_id \\
			recommendationservice2-0 & ProductCatalogService/ListProducts & 29887267 \textmu\textrm{s} & fb9693f175e5b84f\\
		\end{tabular}
		
		\hrule
		\vspace{0.2cm}
		
		\noindent\textbf{[Tool Call]} \texttt{\seqsplit{search\_fluctuating\_metrics(service\_name="recommendationservice", timestamp="1647753157852")}}
		
		\noindent\textbf{[Tool Response]}
		\begin{quote}
			\begin{tabular}{|l|c|c|c|}
				\hline
				\textbf{Metric} & \textbf{regular\_mean} & \textbf{current\_mean} & \textbf{Change} \\
				\hline
				node-5.recommendationservice2-0.pgfault & 0.675 & \textcolor{red}{1.35} & $\times 2.0$ \\
				node-5.recommendationservice-0.pgfault & 0.075 & \textcolor{orange}{0.15} & $\times 2.0$ \\
				node-5.recommendationservice-2.pgfault & 0.025 & \textcolor{orange}{0.05} & $\times 2.0$ \\
				... & ... & ... & ... \\
				\hline
			\end{tabular}
		\end{quote}
		
		\hrule
		\vspace{0.2cm}
		
		\noindent\textbf{[Tool Call]} \texttt{search\_traces(parent\_span\_id="fb9693f175e5b84f")}
		
		\noindent\textbf{[Tool Response]}
		
		\noindent
		\begin{tabular}{llrr}
			cmdb\_id & opeartion & duration & span\_id \\
			productcatalogservice-0 & ProductCatalogService/ListProducts & 25 \textmu\textrm{s} & 12552d251b74a1a4\\
		\end{tabular}
		
		\hrule
		\vspace{0.2cm}
		
		\noindent\textbf{[Tool Call]}
		\begin{quote}
			\begin{verbatim}
				print_results([
				{service: "recommendationservice"}, {pod: "recommendationservice-0"},
				{service: "productcatalogservice"}, {pod: "productcatalogservice-0"},
				{service: "currencyservice"},...])
			\end{verbatim}
		\end{quote}
		
	\end{tcolorbox}
	\caption{Sample Output of a Real Reasoning Case from ThinkFL}
	\label{fig:diagnosis_conversation}
\end{figure}

\section{Sample Output from ThinkFL}\label{sec:appendix-output}

To illustrate how ThinkFL represents and outputs its reasoning, Figure \ref{fig:diagnosis_conversation} presents a real case—chosen because ThinkFL’s final top-1 prediction is incorrect—to highlight both the transparency and practical utility of its reasoning process. For brevity, we show only the tool-invocation dialogue rather than the internal chain-of-thought. As the figure shows, ThinkFL iteratively invokes diagnostic tools, traverses the trace hierarchy, queries metrics for suspected components, and produces a ranked list of candidate root causes with supporting evidence. Although ThinkFL incorrectly ranks RecommendationService above ProductCatalogService, the true root cause is still evident in the intermediate steps: ThinkFL reaches the critical spans, inspects the associated metrics, and includes ProductCatalogService in the final candidate set, demonstrating that the reasoning path remains informative even when the top-1 prediction is wrong.

\section{Systematic Review Process}\label{sec:appendix-survey}

In this section, we present our systematic literature review process and findings on LLM-based failure localization for microservice, including the search strategy and scope, the exclusion criteria, and a summary of the publications that satisfied these criteria.

\subsection{Search strategy}

To manage the vast body of literature efficiently, we limit our search to three databases\footnote{https://ieeexplore.ieee.org/, https://dl.acm.org/, https://arxiv.org/}. We select IEEE Xplore and the ACM Digital Library, as they are frequently cited in related literature surveys and cover a broad range of applied research fields. Finally, we include arXiv, given its prominence as the leading pre-print platform in the field of computer science.

\begin{figure}[h]
	\centering
	\includegraphics[width=\textwidth]{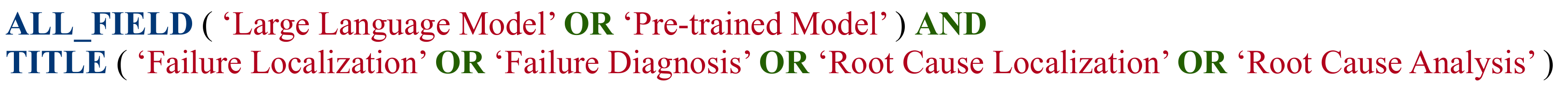}
	\caption{Search Strategy Utilized to Identify Studies on LLM-based Failure Localization}
	\label{fig: search}
\end{figure}

Based on our preliminary exploration, we derived relevant keywords for searching the literature on LLM-based failure localization and restricted our search to papers matching the search string shown in Figure~\ref{fig: search}. To ensure comprehensive coverage, we also included broader keywords such as failure diagnosis and root cause analysis. Subsequently, we manually filter out works that fall outside the scope of this paper using exclusion criteria.

\subsection{Exclusion criteria}

To ensure relevance, we first screen papers by title and abstract; when unclear, we review the full text. We exclude papers that meet any of the following criteria:

\begin{itemize} 
	\item \textbf{EC1:} The study uses a model with a parameter size smaller than 1 billion.
	\item \textbf{EC2:} The study is outside the scope of failure localization.
	\item \textbf{EC3:} The study is unrelated to microservice systems.
	\item \textbf{EC4:} The study only presents conceptual ideas without experimental results.
\end{itemize}

Each record is assessed by a primary reviewer, with non-trivial cases discussed with at least one secondary reviewer. The final list of selected papers is validated by all co-authors. These criteria reflect our goal of focusing on practical, LLM-based failure localization research. EC1 excludes studies using smaller models—particularly BERT variants—which are not generally regarded as LLMs. Because our search strategy is intentionally broad, we manually filter out papers unrelated to failure localization, such as those addressing only failure categorization or program repair. EC3 ensures that selected studies center on microservice systems, excluding work from other domains (e.g., robotics or mechanical devices). Finally, EC4 removes low-quality studies that lack experimental evidence.

\begin{figure}[h]
	\centering
	\includegraphics[width=\textwidth]{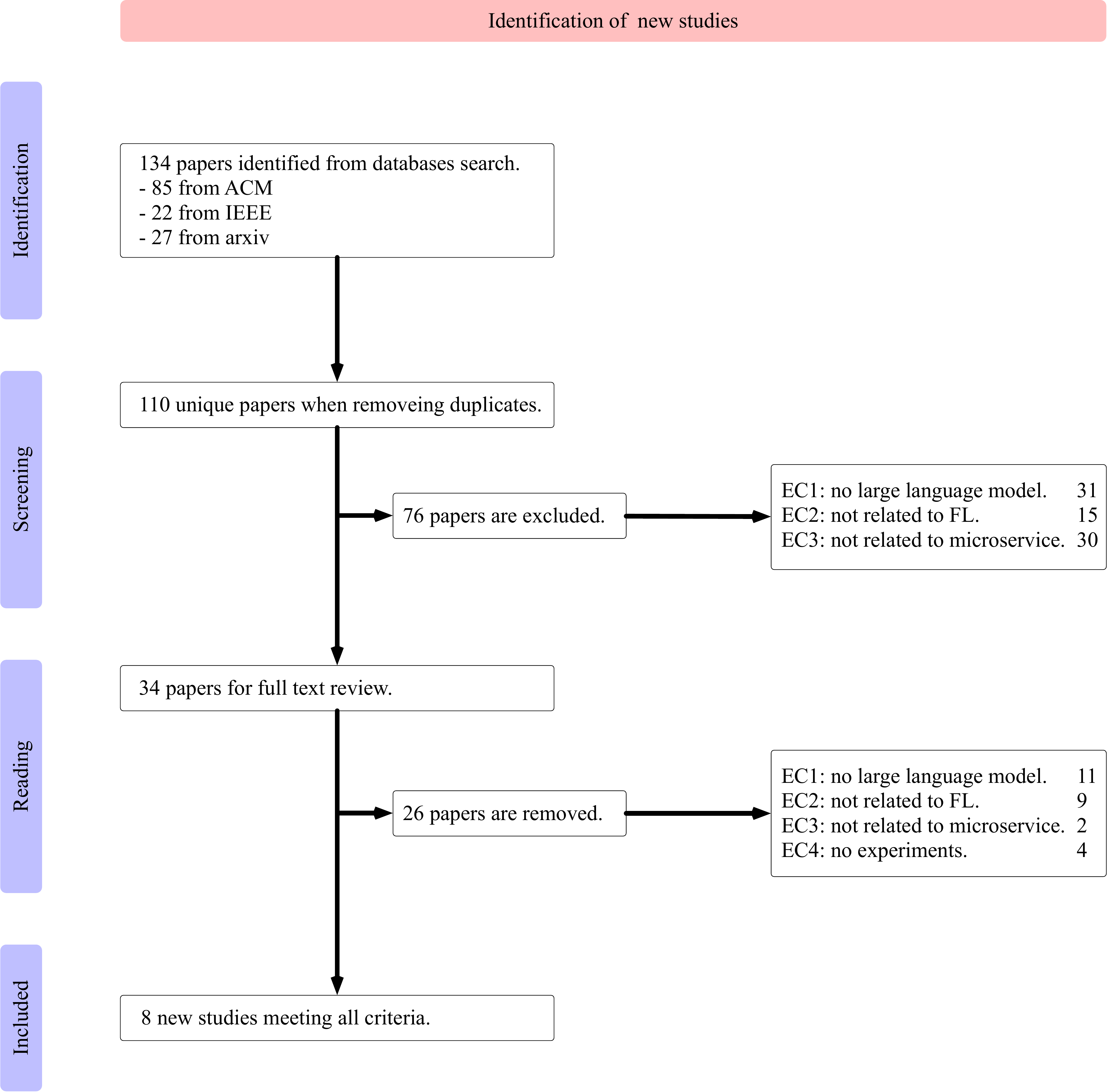}
	\caption{Overview of Paper Selection Procedure}
	\label{fig: selected-publications}
\end{figure}

\subsection{Overview of selected publications}

We then present an overview of the workflow for selecting the papers and summarize the corresponding results. As shown in Figure~\ref{fig: selected-publications}, we initially retrieve 134 papers from three databases. After removing duplicates, 110 papers remain. During the screening phase, we exclude 76 papers based on their titles and abstracts using criteria EC1-EC3. The remaining 34 papers are downloaded for detailed reading, where we find that 26 papers do not align with the scope of our study. Ultimately, 8 new studies satisfy our predefined criteria and are selected for further analysis.

Among the selected papers, six focus directly on failure localization. mABC~\cite{zhang2024mabc} integrates a dependency query engine with a metric explorer to surface suspicious services. RCAgent~\cite{wang2024rcagent} invokes log- and trace-analysis tools to construct structured diagnostic evidence. Flow-of-Action~\cite{pei2025flow} coordinates multiple tools through an SOP-style workflow. COCA~\cite{li2025coca} extracts code snippets and execution paths for localization. TAMO~\cite{wang2025tamo} aligns multi-modal operational signals, and KnowledgeMind~\cite{ren2025multi} pairs metric/log/trace tools with Monte Carlo Tree Search for guided exploration. Two additional works aim to directly generate full root-cause analysis reports, where failure localization constitutes only one component. GALA~\cite{tian2025gala} combines statistical causal inference with LLM-guided iterative reasoning to produce root-cause rankings, incident summaries, and recommended actions. SynergyRCA~\cite{xiang2025simplifying} leverages an LLM to identify key resources and retrieves structured causal context from prebuilt StateGraph and MetaGraph representations to support end-to-end RCA.

\end{document}